\documentclass{aa}  
\usepackage{graphicx}
\usepackage{txfonts}
\begin{document}
   \title{The reversal of the star formation-density relation in the distant universe}

%   \subtitle{}

   \author{
D.~Elbaz\inst{1,2} \and
E.~Daddi\inst{1,2} \and
Damien Le Borgne\inst{1,2} \and
Mark Dickinson\inst{3} \and
Dave M. Alexander\inst{4} \and
Ranga-Ram Chary\inst{5} \and
Jean-Luc Starck\inst{1} \and
William Nielsen Brandt\inst{6} \and
Manfred Kitzbichler\inst{7} \and
Emily MacDonald\inst{3} \and
Mario Nonino\inst{8} \and
Paola Popesso\inst{9} \and
Daniel Stern\inst{10} \and
Eros Vanzella\inst{8}
}

   \offprints{D. Elbaz}

   \institute{CEAÐ-Saclay, DSM/DAPNIA/Service d'Astrophysique, 91191 Gif-sur-Yvette Cedex, France\\
          \email{delbaz@cea.fr}
         \and
AIM-Unit\'e Mixte de Recherche CEA-CNRS (\#7158)- Universit\'e Paris VII
         \and
National Optical Astronomy Observatory, 950 North Cherry Street, Tucson, AZ 85719, USA
         \and
Department of Physics, Durham University, South Road, Durham, DH13LE, UK
         \and
Spitzer Science Center, California Institute of Technology, Pasadena, CA 91125, USA
         \and
Department of Astronomy and Astrophysics, The Pennsylvania State University, 525 Davey Lab, University Park, Pennsylvania, PA 16802, USA
         \and
Max-Planck Institute of Astrophysics, Karl-Schwarzschild Str. 1, PO Box 1317, D-85748 Garching, Germany
         \and
INAF - Osservatorio Astronomico di Trieste, via G.B. Tiepolo 11, 40131 Trieste, Italy
         \and
European Southern Observatory, Karl Schwarzschild Strasse 2, Garching bei Muenchen, 85748, Germany
	\and
Jet Propulsion Laboratory, California Institute of Technology, Pasadena, CA 91109, USA
             }

   \date{Received March 22, 2007; accepted April 5, 2007}

% \abstract{}{}{}{}{} 
% 5 {} token are mandatory
 
  \abstract
   {}
   {We study the relationship between the local environment of galaxies and their star formation rate (SFR) in the Great Observatories Origins Deep Survey, GOODS, at $z\sim$1.
   }
   {We use ultradeep imaging at 24\,$\mu$m with the MIPS camera onboard ${\it Spitzer}$ to determine the contribution of obscured light to the SFR of galaxies over the redshift range 0.8$\leq z \leq$1.2. Accurate galaxy densities are measured
    thanks to the large sample of $\sim$1200 spectroscopic redshifts with high ($\sim$70\,\%) spectroscopic completeness. 
    Morphology and stellar masses are derived from deep HST-ACS imaging, supplemented by ground based imaging programs and photometry from the IRAC camera onboard ${\it Spitzer}$. 
  }
   {We show that the star formation--density relation observed locally was reversed at $z\sim$ 1: the average SFR of an individual galaxy increased with local galaxy density when the universe was less than half its present age.  Hierarchical galaxy formation models (simulated lightcones from the Millennium model)
predicted such a reversal to occur only at earlier epochs (z$>$2) and at a lower level. We present a remarkable structure at $z\sim$ 1.016, containing X-ray traced galaxy concentrations, which will eventually merge into a Virgo-like cluster. This structure illustrates how the individual SFR of galaxies increases with density and shows that it is the $\sim$1--2 Mpc scale that affects most the star formation in galaxies at $z\sim$ 1.
The SFR of $z\sim$ 1 galaxies is found to correlate with stellar mass suggesting that mass plays a role in the observed star formation--density trend. However the specific SFR (=SFR/M$_{\star}$) decreases with stellar mass while it increases with galaxy density, which implies that the environment does directly affect the star formation activity of galaxies. Major mergers do not appear to be the unique or even major cause for this effect since nearly half (46\,\%) of the luminous infrared galaxies (LIRGs) at $z\sim$ 1 present the HST-ACS morphology of spirals, while only a third present a clear signature of major mergers. The remaining galaxies are divided into compact (9\,\%) and irregular (14\,\%) galaxies. Moreover, the specific SFR of major mergers is only marginally stronger than that of spirals. }
{These findings constrain the influence of the growth of large-scale structures on the star formation history of galaxies. Reproducing the SFR--density relation at $z\sim$ 1 is a new challenge for models, requiring a correct balance between mass assembly through mergers and in-situ star formation at early epochs.}

   \maketitle
%
%________________________________________________________________
\section{Introduction}
The activity of star formation in a present-day galaxy is strongly related to the local density of galaxies surrounding it and to its mass: massive galaxies both lie in denser environments and are populated by redder, colder stars than less massives ones (Kauffmann et al. 2004). The stellar abundance ratios and age estimates of present-day red galaxies indicates that they formed their stars on shorter timescales (Thomas et al. 2005), which suggests that the star formation rate (SFR) density was larger in dense environments in the past. The epoch at which the environment affected the activity of galaxies is a key issue for modern cosmology. Hence, measuring the dependence of the average instantaneous SFR per galaxy with environment at early epochs, when the  SFR density of the Universe was much larger than today, would provide a strong test for galaxy formation models.

The recent advent of complete spectroscopic redshift surveys in the local universe with the Sloan Digital Sky Survey (SDSS) and the Two Degree Field Galaxy Redshift Survey (2dFGRS) have made it possible to extend the study of galaxy properties down to very low galaxy densities, ranging from clusters to field galaxies. With the discovery that galaxy colors, H$\alpha$ emission, stellar mass and morphology varied over this extended range of environments, it appeared that the physical mechanisms which related galaxies with their environment was more complex than initially thought and could not be reduced to the mere ram pressure stripping from the intra-cluster medium of massive galaxy clusters, as past studies were suggesting being limited to the comparison between field and cluster galaxies (see discussion in Balogh et al. 2004).

The next step to understand how and if the environment affects galaxies, after having extended its measurement to low densities, is to measure galaxy properties as a function of density and redshift. The Butcher-Oemler effect (Butcher \& Oemler 1978, 1984) was the first  attempt in doing so with the finding that the fraction of blue galaxies detected in clusters was increasing with redshift. However, the interpretation of this effect has become controversial since it was found that the average star formation history of a galaxy presents a rapid decline since $z\sim$ 1 (Lilly et al. 1996, Madau et al. 1996, Chary \& Elbaz 2001, Le Floc'h et al. 2005, Hopkins \& Beacom 2006). If field galaxies were forming more stars in the past, hence were bluer, then the infall of field galaxies in clusters could alone explain the Butcher-Oemler effect. Using a more direct determination of the instantaneous star formation rate (SFR) of galaxies, the mid-infrared (mid-IR), Fadda et al. (2002) found that the evolution of field galaxies alone was probably not enough to explain the excess of star formation activity in Abell 1689 ($z=0.2$). However, cluster galaxies only represent 5 to 10\,\% of present-day galaxies and are therefore not best suited to understand the origin of the local galaxy-density relationship found by the 2dF and SDSS. 

The modern way to look at the evolution of the optical color of galaxies is to follow the build-up through time of their bimodal distribution in color-magnitude diagrams, typically U-B versus B, between red-dead and blue-active galaxies. Recent studies have demonstrated that this bimodality was already in place at $z\sim$1, with red galaxies preferentially located in the highest density regions, as it is also the case locally (Cooper et al. 2006, Cucciati et al. 2006). Coupled with evidence that the average strength of the oxygen emission line, as traced by its equivalent width (EW[OII]), decreases with galaxy density (Cooper et al. 2006), this behaviour was interpreted as evidence that at $z\sim$1 star formation was more efficiently quenched by the environment in massive galaxies, i.e. a downsizing of quenching. However, the color of a galaxy is not straightforwardly linked with its instantaneous SFR but instead reflects a combination of stellar mass, star formation history over the lifetime of the galaxy and SFR. Moreover, the EW[OII] itself is less directly related to the SFR than to the birthrate parameter of a galaxy (b=SFR/$<$SFR$>$, where $<$SFR$>$ is the past average SFR history), and is very sensitive to dust extinction and metallicity (Jansen, Franx \& Fabricant  2001).

The bimodality of galaxies was initially measured through the morphology-density segregation, with the fraction of spirals decreasing from 80 to 10\,\% going from rich clusters to the field (Dressler 1980). But its evolution with redshift is harder to quantifiy not only because of instrumental difficulties, which have become less of a problem with the HST, nor due to k-correction, but because the morphology of galaxies evolves so much with redshift, with a rapid increase of the fraction of irregular galaxies, that the classical morphology criteria, such as the bulge to disk ratio, cannot easily be translated to the distant universe.

An optimum strategy to understand the star formation--density relation is therefore to directly measure the instantaneous SFR of galaxies as a function of time (redshift) and environment. There are two major difficulties associated to this strategy. First, it is necessary to combine direct and re-radiated O and B stellar light
to derive the actual star formation rate (SFR) of galaxies which requires 
to  take into account both UV (dominating at low SFR) and IR (dominating the
  bolometric luminosity at high SFR) photons. This is particularly relevant since the contribution of dust reprocessed starlight to the density of SFR per unit comoving volume (SFRD) increases with redshift together with the role of luminous (LIRGs, 11$\leq log_{10}\left(L_{\rm IR}\right)\leq 12$) and ultra-luminous (ULIRGs, 12$\leq log_{10}\left(L_{\rm IR}\right)\leq 13$) IR galaxies (Chary \& Elbaz 2001, Le Floc'h et al. 2005). Second, high spectroscopic completeness is required to measure the local galaxy density, since the uncertainty on photometric
  redshifts converts into a comoving size which is larger than the
  transversal size of the GOODS fields (10$\times$16 comoving Mpc at
  $z\sim$ 1).

Here, we overcome these limitations thanks to the deepest existing
mid-infrared survey at 24\,$\mu$m in two fields of 10$\arcmin\times$16$\arcmin$, with some of the largest spectroscopic completeness. An important point to note is that this size is well suited for studying the effect of the local environment on the star formation activity of galaxies at the $\sim$1 Mpc scale. Investigations using the Sloan Digital Sky Survey (SDSS) indeed suggest that this is the physical scale that mostly affects the SFR of galaxies (Blanton et al. 2006). 

The aim of the present paper is to study the relationship between the SFR and local galaxy density
 down to the Milky-Way activity at $z\sim$ 1,
where galaxy densities are measured at the mega-parsec scale. Our results
at $z$=0.8--1.2 are compared to the local universe, with the SDSS data,
and to numerical simulations (Millennium Run).

The samples are presented in Sect.~\ref{SEC:samples} followed by the description of the spectroscopic completeness and large-scale structure content of the GOODS fields (Sect.~\ref{SEC:zspec}). We then present how the local galaxy density and SFR are measured (Sect.~\ref{SEC:definition}). The main result of the paper, the reversal of the star formation-density relation at $z\sim$1, is presented in Sect.~\ref{SEC:result}, followed by a discussion of the systematics in the determination of the average SFR and local galaxy density in Sect.~\ref{SEC:systematics} (the potential contamination by active galactic nuclei, AGNs, Sect.~\ref{SEC:AGNs}, and selection effects due to the spectroscopic incompleteness, Sect.~\ref{SEC:completeness}). We then compare this result with previous findings on the influence of the environment on the bimodality of galaxies (Sect.~\ref{SEC:bimodality}). 

Sect.~\ref{SEC:discussion} is devoted to the discussion of the origin of the reversal of the star formation--density relation. The connection with large-scale structure formation is discussed in Sect.~\ref{SEC:cluster}, where we present the discovery of a candidate proto-Virgo cluster at $z\sim$1.016. We then present a correlation between the stellar mass and SFR of $z\sim$1 galaxies in Sect.~\ref{SEC:SFRMstar} which could be responsible for the reversal due to the preferential presence of massive galaxies in dense regions. To disentangle the two possibilities (external environment effect versus internal stellar mass), we study the specific SFR (=SFR/M$_{\star}$) of the $z\sim$1 galaxies as a function of stellar mass, galaxy density (Sect.~\ref{SEC:SSFR}) and morphology (Sect.~\ref{SEC:mergers}). We classify LIRGs at $z\sim$1 in four morphological types and discuss the role of major mergers as a potential driver of the influence of the galaxy density on the star formation activity and on the triggering of the LIRG phase in a galaxy (Sect.~\ref{SEC:mergers}).

We will use in the following a cosmology of: H$_{0}$= 70 km s$^{-1}$
Mpc$^{-1}$, $\Lambda$= 0.7, $\Omega_{\rm m}$=0.3.

\section{Definition of the samples}
\label{SEC:samples}

\subsection{Description of the GOODS sample}
The Great Observatories Origins Deep Survey (GOODS, see Dickinson \& Giavalisco 2002) is a {\it Spitzer} Legacy Program (PI M.Dickinson) and an Hubble Space Telescope (HST) Treasury Program (PI M.Giavalisco) consisting of two fields from the two hemispheres, GOODS-North and GOODS-South (hereafter GOODS-N and GOODS-S). The GOODS-N field is centered on the Hubble Deep Field North (HDFN; 12h36m49.4s, +62$^o$12$\arcmin$58.0$\arcsec$) and GOODS-S is contained within the Chandra Deep Field South (CDFS; 3h32m28.0s, -27$^o$48$\arcmin$30.0$\arcsec$). The size of each field is 10$\arcmin\times16\arcmin$. The wide separation of the two GOODS fields in both hemispheres will allow us to test our results against the effects of cosmic variance. For a detailed description of all the ancillary data existing in the GOODS fields we refer to the GOODS public webpage (http://www.stsci.edu/science/goods/).

Both fields have been the subject of some of the largest, deepest, and most complete spectroscopic campaigns to date as well as multiwavelength ground-based imaging. Deep imaging with the Advanced Camera for Surveys (ACS) onboard the HST has been used to obtain photometry and morphological information, supplemented by a variety of optical and near-IR ground based deep imaging programs (Giavalisco et al. 2004). The two GOODS fields were imaged in the four ACS bands - F435W (B), F606W (V), F775W (i), and F850LP (z) - down to the 5-$\sigma$ sensitivity limits in the AB system of 27.9, 28.2, 27.5, 27.4 respectively. In order to insure a good spectroscopic completeness, we will restrict the GOODS sample to z$_{\rm AB}$=23.5, which is equivalent to M$_B=$-20 at $z=$1 (see Sect.~\ref{SEC:zspec}).

\subsubsection{{\it Spitzer}-IRAC data}
We used the photometry at 3.6, 4.5\,$\mu$m (down to AB = 24.5, 24.5;  equivalent to 0.6--1.2\,$\mu$Jy) obtained with the {\it Spitzer} (Werner et al. 2004) infrared array camera (IRAC; Fazio et al. 2004) to derive stellar masses and photometric redshifts using PEGASE.2 (Fioc \& Rocca-Volmerange 1997, 1999). The 5.8 and 8.0\,$\mu$m passbands (AB = 23.8, 23.7) were not used here because for distant galaxies they combine stellar and dust emission which are difficult to separate and may therefore provide wrong stellar masses estimates. 

\subsubsection{{\it Spitzer}-MIPS data 24\,$\mu$m}
Ultradeep imaging at 24\,$\mu$m was obtained with the MIPS camera (Multiband Imaging Photometer for {\it Spitzer}; Rieke et al. 2004) onboard {\it Spitzer} down to a 5-$\sigma$ point source sensitivity limit of 25\,$\mu$Jy in both GOODS fields (Chary 2006). This is a crucial advantage of the present study as it allows us to determine the contribution of dust obscured light to the star formation rate (SFR) of galaxies down to very low SFR. The {\it Spitzer} beam size at 24\,$\mu$m is 5.7$\arcsec$ FWHM which raises the issue of source confusion. The surface density of sources is 17 arcmin$^{-2}$ in GOODS-N and 24 arcmin$^{-2}$ in GOODS-S, hence there is a $\sim$20\,\% probability that two GOODS sources are closer than the FWHM (cf Dole, Lagache \& Puget 2003). However, using a prior position based point source fitting algorithm, we have been able to extract sources with a completeness of 84\,\% at 24\,$\mu$Jy. Indeed every MIPS source is detected in the deep IRAC--3.6\,$\mu$m observations. Note that due to the accurate pointing of {\it Spitzer}, the 24\,$\mu$m images can be aligned with the IRAC ones with a precision of $\sim$0.2$\arcsec$ $rms$.

\subsection{Millennium lightcones}
The series of simulated lightcones of the Universe that are used in this work was produced by Kitzbichler \& White (2007), who applied the semi-analytical model of Croton et al. (2006) to the Millennium Run simulation (Springel et al. 2005). We applied the same selection criterion that we used for GOODS (M$_{\rm B}\leq$ -20) to galaxies in the simulations. We computed the local galaxy density, $\Sigma$ (described in Sect.~\ref{SEC:galdens}), with the same technique as for GOODS. We find that the average comoving density of galaxies inside 0.8$\leq z \leq$1.2 in the "Millennium model" matches closely that in the GOODS fields. We use the Millennium lightcones to estimate the star formation-density relation in various redshifts intervals for comparison with the observed GOODS dataset and also to investigate the robustness of our result against spectroscopic incompleteness and cosmic variance associated to the limited size of the survey.

\subsection{The Sloan Digital Sky Survey dataset}
%__________________________________________________________________
   \begin{figure}
   \centering
   \includegraphics[width=8cm]{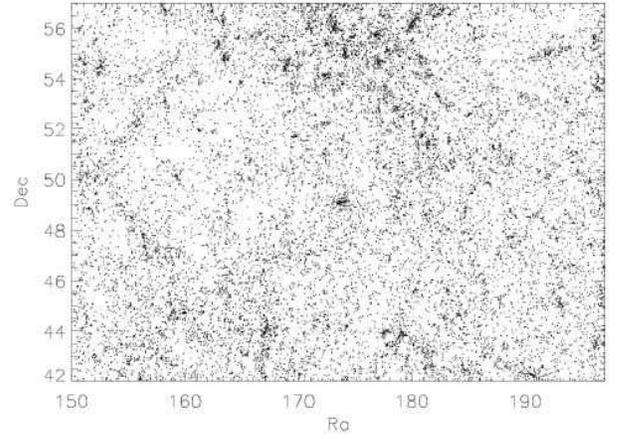}
      \caption{Spatial distribution of the galaxies in the region selected for the study of the link between the environment and SFR in the SDSS (0.015$\geq~z~\geq$0.1).
              }
         \label{FIG:RADEC}
   \end{figure}
%__________________________________________________________________
%__________________________________________________________________
   \begin{figure}
   \centering
   \includegraphics[width=8cm]{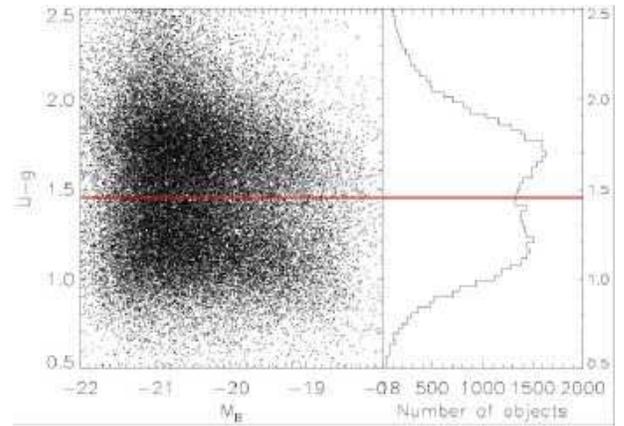}
      \caption{Rest-frame colour-magnitude diagram for the SDSS galaxies. The $g$-band is centered at 4825\,\AA, hence close to the B-band (4400\,\AA) often used to separate red and blue galaxies.
              }
         \label{FIG:SDSScolma}
   \end{figure}
%__________________________________________________________________
%__________________________________________________________________
   \begin{figure}
   \centering
   \includegraphics[width=8cm]{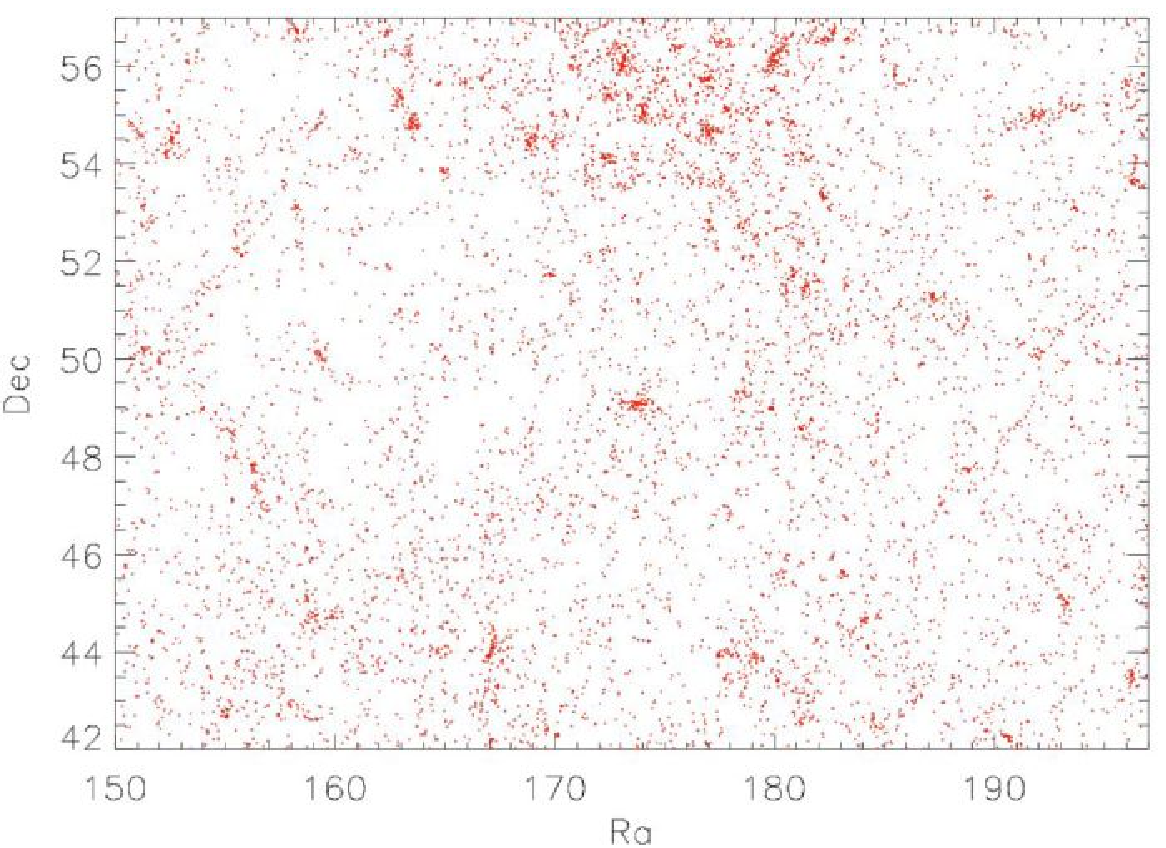}
   \includegraphics[width=8cm]{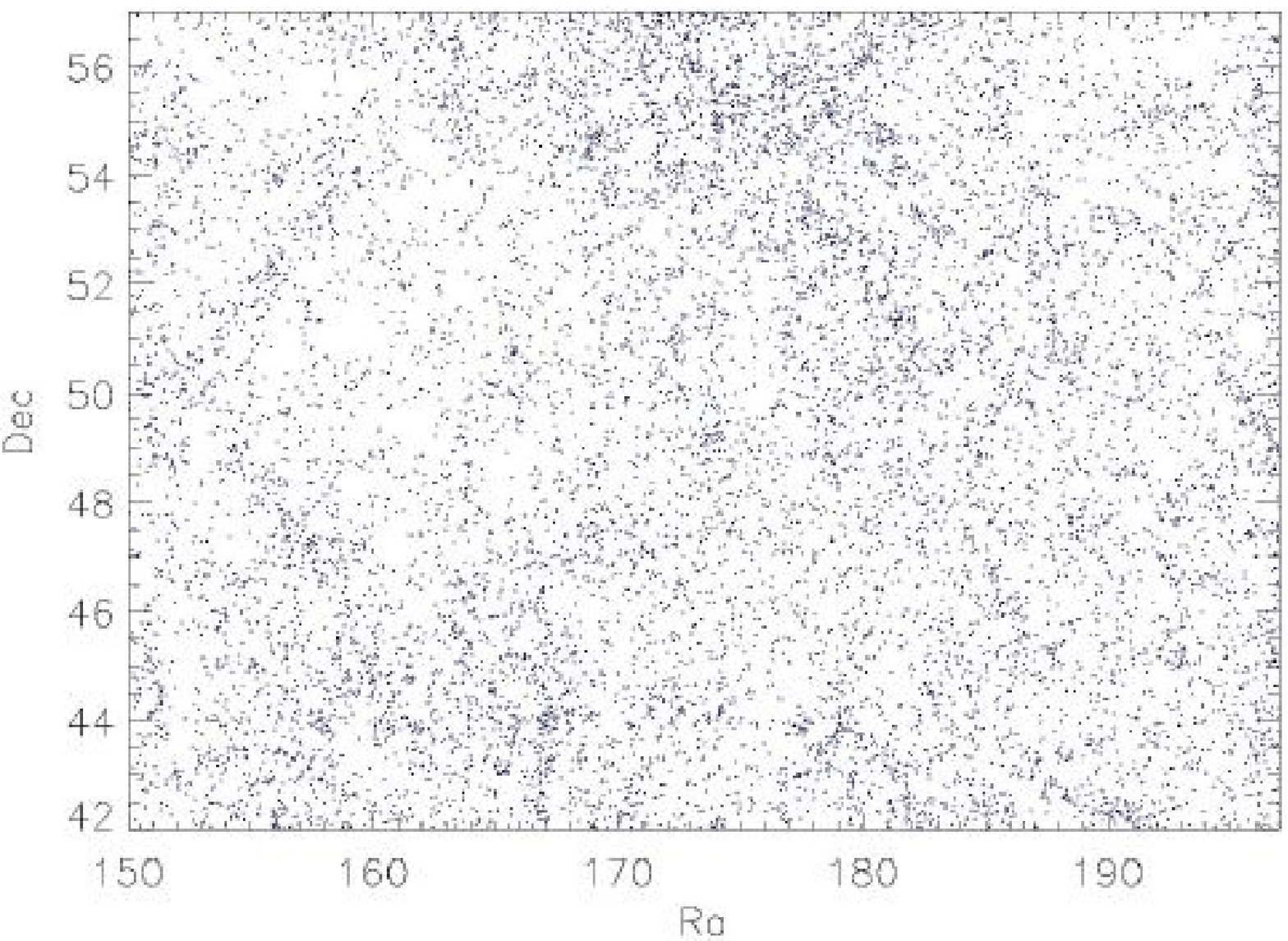}
      \caption{Spatial distribution of the red (top) and blue (bottom) galaxies in the region selected for the study of the link between the environment and SFR in the SDSS (0.015$\geq z \geq$0.1). Red galaxies clearly follow the spatial distribution of the densest concentrations of galaxies, while blue galaxies present a nearly random spatial distribution..
              }
         \label{FIG:RADECred}
   \end{figure}
%__________________________________________________________________
We have selected the widest region with contiguous spectroscopic coverage that we could find within the Sloan Digital Sky Survey (SDSS) to study the relation between the SFR and local environment in the local universe. This region of 705 square degrees centered at (11h40m, 50$\degr$) presents a fair representation of various environments including galaxy clusters which can be seen in the Fig.~\ref{FIG:RADEC}. The local comparison sample that we have produced using the SDSS dataset was limited to redshifts between $z$=0.04--0.1, where the SDSS spectroscopy is fairly complete to M$_{\rm B}\leq$-20 (see Berlind et al. 2006) with 19590 galaxies (Data Release 4 of the SDSS). The minimum redshift of $z=$0.04 was chosen to avoid missing light from closeby galaxies due to aperture effects (see Kewley, Jansen \& Geller 2005). We applied the same selection criterion that we used for GOODS in the SDSS (M$_{\rm B}\leq$-20). 
 
The SFR and M$_{\star}$ for the Data Release 4 were obtained from the internet database at http://www.mpa-garching.mpg.de/SDSS/DR4/. The SFR were derived by Brinchmann et al. (2004) from the H$_{\alpha}$ emission line corrected for extinction using the H$_{\alpha}$/H$_{\beta}$ ratio. The derivation of the intrinsic SFR of local SDSS galaxies is in fact slightly more complex than just measuring the Balmer decrement and deriving the associated dust attenuation, since the authors were able to correct for the underlying photospheric absorption line depending on the spectral resolution and signal-to-noise ratio on the continuum of the spectra and needed to correct for fiber aperture effects (for a detailed description of the technique see Brinchmann et al. 2004). The stellar masses have been derived by Kauffmann et al. (2003) using a library of Monte Carlo realizations of different star formation histories (including starbursts of varying strengths and a range of metallicities) generated using the Bruzual and Charlot (2003) code (with a 3\,\AA\,FWHM resolution) using a bayesian statistics to derive the probability distribution functions of the parameters involved in the stellar mass derivation for the galaxies. Both the SFR and M$_{\star}$ were computed for a Kroupa IMF that we converted to a Salpeter IMF, by dividing them by a factor of 0.7. 

If we separate red and blue galaxies following their bimodal distribution in the U-g versus M$_{\rm B}$ color-magnitude diagram (Fig.~\ref{FIG:SDSScolma}), we find as expected that red galaxies clearly inhabit the densest regions of the local universe (Fig.~\ref{FIG:RADECred}-top) while blue galaxies are more homogeneously distributed (Fig.~\ref{FIG:RADECred}-bottom). The separation of SDSS galaxies into blue and red galaxies will be used in Sect.\ref{SEC:SFRMstar} where we discuss the relation between stellar mass and SFR.

\section{Spectroscopic completeness and large-scale structure content of the GOODS fields}
\label{SEC:zspec}
%__________________________________________________________________
   \begin{figure}
   \centering
   \includegraphics[width=9cm]{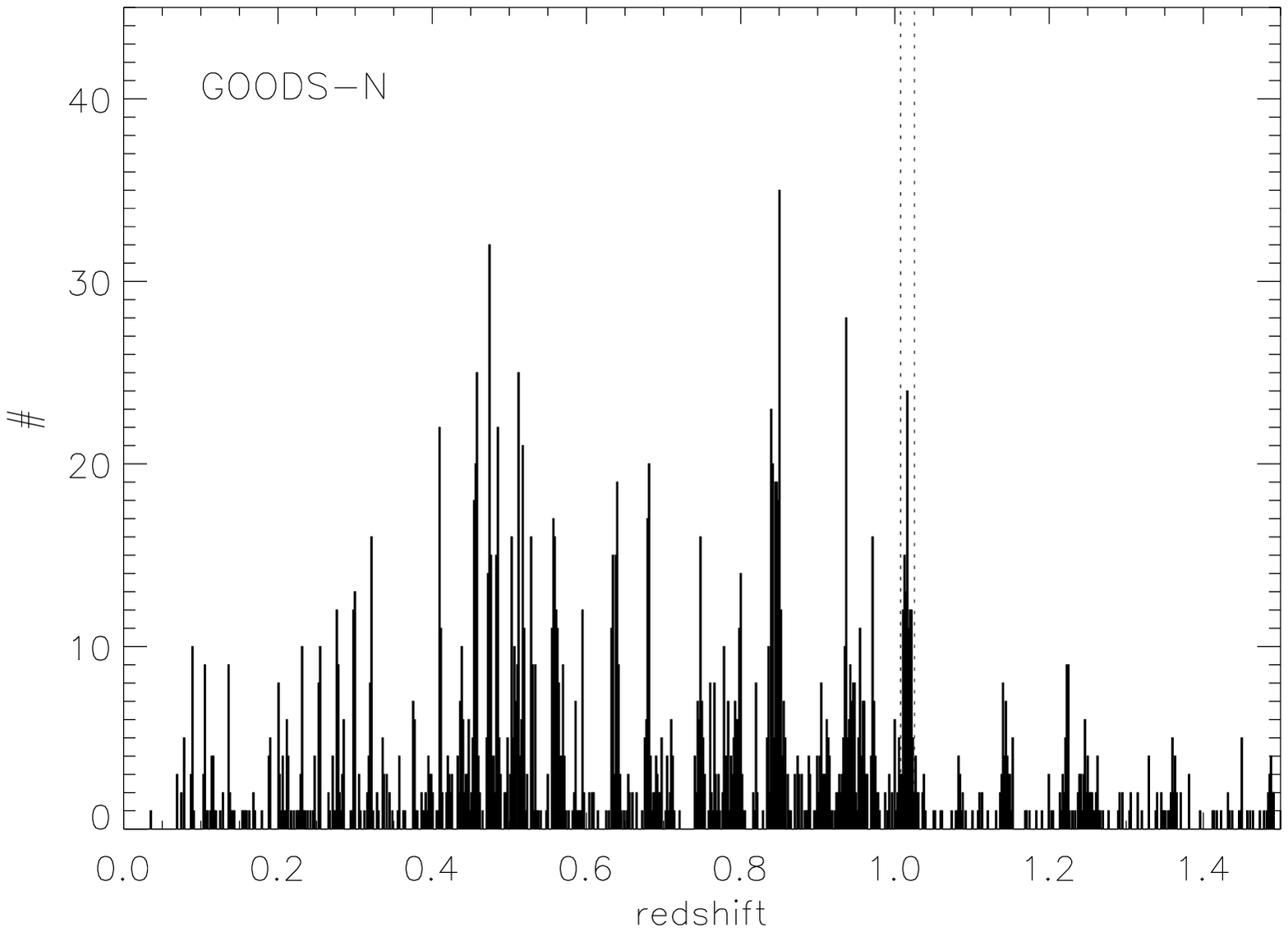}
    \includegraphics[width=9cm]{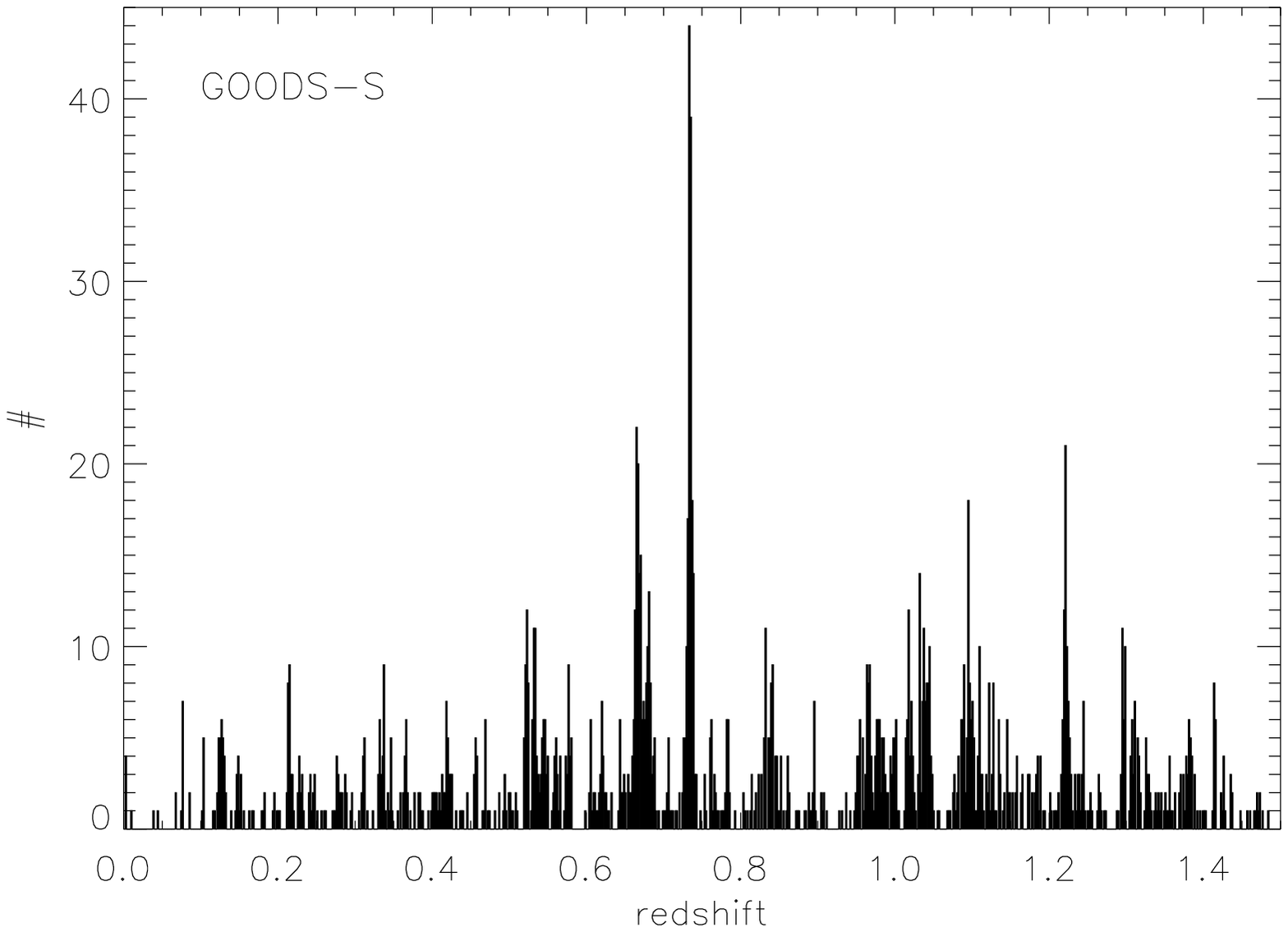}
     \caption{Distribution of the spectroscopic redshifts in the GOODS-N (top; 2108 galaxies) and GOODS-S (bottom; 1781 galaxies) field inside 0$\leq$z$\leq$1.5 (redshift step: $\Delta$z=0.02). The cosmic variance is clearly visible on these two figures.
              }
         \label{FIG:hiszGN}
   \end{figure}
%__________________________________________________________________
%__________________________________________________________________
   \begin{figure}
   \centering
   \includegraphics[width=8cm]{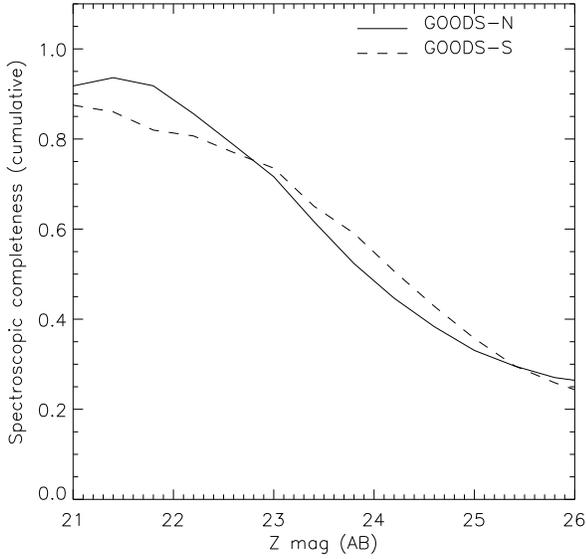}
      \caption{Spectroscopic completeness in the GOODS fields as a function of z-band magnitude (AB) for galaxies with a photometric redshift within $z_{\rm phot}$=0.8-1.2. The completeness is cumulative, i.e. the lines represent the fraction of galaxies brighter than z$_{\rm AB}$ for which a spectroscopic redshift was measured.
              }
         \label{FIG:zABcompleteness}
   \end{figure}
%__________________________________________________________________

%__________________________________________________________________
   \begin{figure}
   \centering
   \includegraphics[width=8cm]{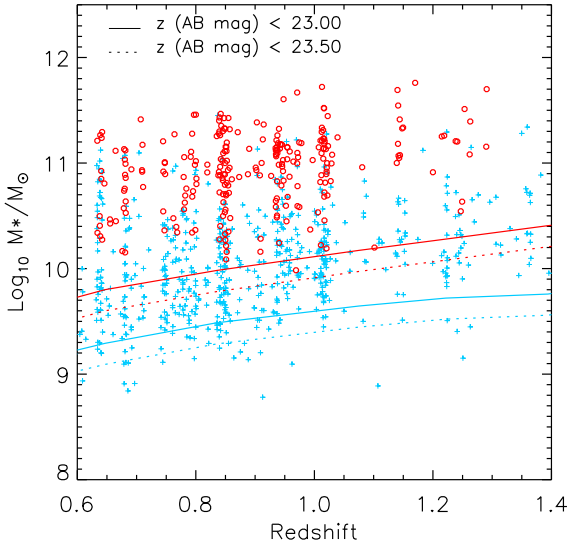}
      \caption{Stellar mass as a function of redshift for two limiting magnitudes z$_{\rm AB}$ (23-plain, 23.5-dashed lines) for an instantaneous burst at $z$=10 (red) and a continuously star forming galaxy (blue).  A z$_{\rm AB}$ cutoff of 23.5 for 0.8$\leq$z$\leq$1.2 as in the present study will therefore select all galaxies with M$_{\star}\geq~10^{10}$ M$_{\sun}$.
              }
         \label{FIG:Mstarcompleteness}
   \end{figure}
%__________________________________________________________________
\subsection{Spectroscopic completeness}
A large number of spectroscopic redshifts have been measured for galaxies in both GOODS-N (2376 galaxies; compilation of the Caltech Faint Galaxy Redshift Survey- CFGRS, Cohen et al. 2000-, the Team Keck Treasury Redshift Survey, Wirth et al. 2004, Cowie et al. 2004; Stern et al. in prep) and GOODS-S (2547 galaxies ; Vanzella et al. 2006, Le F\`evre et al. 2004, Mignoli et al. 2005). Their distribution between z=0 and 1.5 is highly clustered in the redshift space with most galaxies lying inside redshift peaks of width $\Delta z\sim$0.02 (Fig.~\ref{FIG:hiszGN}; see also Cohen et al. 2000). The structure at $z=$1.016$\pm$0.009 (highlighted with dashed lines in Fig.~\ref{FIG:hiszGN}-top) corresponds to one of the strongest redshift peaks in GOODS-N.

Because the spectroscopic follow-up of the GOODS fields has been obtained from several campaigns with diverse selection criteria, it is important to quantify their net completeness. This is particularly relevant where we compare the properties of $z\sim$1 galaxies to those observed in the present-day universe and to simulations, where we apply a similar selection criterion.  Spectroscopic redshifts have been obtained for 60\,\% of the galaxies in GOODS-N and 50\,\% in GOODS-S down to z$_{\rm AB}$=23.5. In order to quantify the completeness of the GOODS spectroscopic sample inside the redshift range of interest here (0.8$\leq z \leq$1.2), we used the large number of passbands available in the GOODS fields (UBVRIzJHK,3.6\,$\mu$m,4.5\,$\mu$m) to derive photometric redshifts with an accuracy of $\Delta z$/(1+$z$)=0.05--1 using Z-PEG (Le Borgne \& Rocca-Volmerange 2002). This allowed us to construct the reference sample of galaxies potentially located within 0.8$\leq z \leq$1.2, where the photometric redshifts are very robust. When compared to this reference sample, we find that the fraction of galaxies with a spectroscopic redshift inside 0.8$\leq z_{\rm phot} \leq$1.2, is $\sim$75\,\% for both fields down to z$_{\rm AB}$=23 and 67\,\% (and 60\,\%) for GOODS-N (and GOODS-S) galaxies brighter than z$_{\rm AB}$=23.5 (see Fig.~\ref{FIG:zABcompleteness}). In the following, we will use z$_{\rm AB}$=23.5 as the magnitude limit of our sample and not consider fainter galaxies, for consistency with the comparison samples. This corresponds to the limit M$_B\leq$-20 at $z=$1.

This selection criterion can be converted into an equivalent stellar mass for different star formation histories of an individual galaxy. Two extreme cases, computed with PEGASE.2 (Fioc \& Rocca-Volmerange 1997, 1999), are presented in Fig.~\ref{FIG:Mstarcompleteness}: a galaxy observed at $z$=1 after an instantaneous burst at $z$=10 without any other star formation (red line) and a continuously star forming galaxy (blue line). The magnitude cutoff that we have applied is equivalent to a stellar mass cutoff of M$_{\star}\geq 10^{10}$ M$_{\sun}$ and M$_{\star}\geq 10^{9}$ M$_{\sun}$ in the two extreme cases in the 0.8$\leq z \leq$1.2 domain. Indeed, we find that the red and blue galaxies in GOODS (separated from the bimodality in the color-magnitude diagram presented in Sect.~\ref{SEC:bimodality}) fall above the dashed red and blue lines, respectively.

\subsection{Cosmic volume associated to GOODS and large-scale structure content}
Over the redshift range 0.8$\leq z\leq$1.2, each field spans a volume of 1.4$\times$10$^5$ Mpc$^3$ and contains of order 1000 galaxies with measured spectroscopic redshifts. However, we note that the volumes surveyed are not large enough to include the rarest and strongest overdensities. We used the 0.5-2 keV luminosity function of X-ray clusters (Fig.9 from Rosati, Borgani \& Norman 2002) to estimate the number of large-scale structures such as groups or clusters expected in a 160$\arcmin^2$ field, such as the GOODS fields studied here, at $z=$ 0.8--1.2 (=1.4$\times$10$^5$ Mpc$^3$). Note that within the error bars there is little, if any, redshift evolution of the cluster space density at L$_X$[0.5-Ð2 keV]$\leq$3$\times$ 10$^{44}$ erg s$^{-1}\sim$ L$_X^{\star}$. 
The field size would have to be of 2-3 degrees on a side to include a Coma sized cluster (L$_X$[0.5-Ð2 keV]=4.6$\times$10$^{44}$ erg s$^{-1}$, Briel, Henry \& Boehringer 1992) within 0.8$\leq z\leq$1.2 and $\sim$50$\arcmin$ to include a Virgo sized cluster  (L$_X$[0.5-Ð2 keV]=3.5$\times$10$^{43}$ erg s$^{-1}$, derived from Boehringer et al. 1994). Hence, the discussion of the role of X-ray emitting large-scale structures on galaxy evolution is out of the scope of the present paper. However, using the same Eq.~\ref{EQ:NbX}, we should statistically find one L$_X$[0.5-Ð2 keV]=2$\times$10$^{42}$ erg s$^{-1}$ group per GOODS field and we do indeed detect such a group at $z\sim$1.016 that we associate to a proto-Virgo cluster (see Sect.~\ref{SEC:cluster}).

   \begin{equation}
   \label{EQ:NbX}
   \begin{array}{l}
N(>L_X,0.8\leq z\leq1.2)= 5.7 \times (L_{43})^{-1} ~ \left[{\rm cluster}/\degr ^{2}\right] \\
{\rm where:~} L_{43}=L_X[0.5-Ð2 {\rm keV}]/10^{43} ~{\rm erg~s^{-1}}
   \end{array}
   \end{equation}

\section{Description of the measurement of the local galaxy density and star formation rate}
\label{SEC:definition}
\subsection{Definition of local galaxy density}
\label{SEC:galdens}
We have computed the local galaxy density, $\Sigma$, by counting all galaxies located inside boxes of comoving size 1.5 Mpc in the right ascension (Ra) and declination (Dec) directions, i.e. approximately 1.5 arcmin at $z\sim$1. The size of the box in the redshift direction was chosen to correspond to $\Delta $v= 3000 km s$^{-1}$ at $z$=1, i.e. $\Delta z$=0.02 ($\Delta $v=$\Delta z$/(1+$z$)). This choice was driven by two requirements : 

-	the need to use a velocity dispersion larger than that due to the internal velocity dispersion inside massive galaxy clusters ($\sim$1000 km s$^{-1}$) and than the spectroscopic redshift errors of $\delta z \sim$2.

-	to avoid splitting redshift peaks in the redshift distribution in order to sample the largest range of local environments for the galaxies. The full width of the peaks is of the order of $\Delta z\sim$0.02, computed using the technique described by Gilli et al. (2003). 

At redshifts different than $z$=1, the length of the boxes in the redshift direction was adjusted to keep the same comoving volume inside the boxes (90 Mpc$^{3}$), i.e. 40 comoving-Mpc.  The projected galaxy density, $\Sigma$, was then computed by dividing the number of sources by (1.5 Mpc)$^2$ and for the volume density, $\rho$,  by 90 Mpc$^3$.
 
We compared two different techniques to compute the average SFR of galaxies as a function of galaxy density, $\Sigma$. In the first technique, we virtually associated a box to every single galaxy and computed its local environment within this box as well as the average SFR of galaxies inside this box. In the second technique, we moved the boxes independently of the positions of the galaxies by a step in Ra and Dec. The difference between the two techniques is negligible. In the following of the paper we will use the first technique.

After having defined an average SFR, $<$SFR$>$, for all galaxies inside the 1.5$\times$1.5$\times$40 Mpc$^3$ boxes, we grouped them in density bins to derive a distribution of $<$SFR$>$ inside a given density range. We then derived from this distribution the typical  $<$SFR$>$ for each density bin, computed as the median of the distribution, with an associated error bar that we computed using a bootstrap technique. If N$_{\rm tot}$ is the number of boxes within the selected density interval, we randomly extracted N$_{\rm box}$ boxes from this sample and computed their median SFR. This process was repeated 500 times. The error bar on the typical  $<$SFR$>$ for a given density bin is equal to the root mean square of the 500 measurements of SFR divided by the square root of N$_{\rm tot}$/N$_{\rm box}$. We checked that the error bar was robust by trying several values for N$_{\rm box}$. In the highest density bins, where the number of galaxies gets small, we used the median of the $<$SFR$>$ in the density range divided by the square root of the number of galaxies.

\subsection{Estimation of the SFR}
\label{SEC:definitionSFR}
While it has been demonstrated that the mid infrared is a robust tracer (at the 40\,\% rms level) of the total IR luminosity (L$_{\rm IR}$=L[8--1000\,$\mu$m]) in the present-day universe (Chary \& Elbaz 2001, Spinoglio et al. 1995), it is to be expected that the mid to far IR correlations were different in the past, when the metal content, grain size distribution, dust geometrical distribution and ionization rate were different. However, in the redshift range of interest here, i.e. around $z\sim$ 1, the deepest existing radio data have shown that the mid IR and radio data were providing consistent Lir determinations, at the 40\,\% rms level, up to $z\sim$ 1.2 in the LIRG regime (Elbaz et al. 2002, Appleton et al. 2004). This is not unexpected since the metal content of $z\sim$ 1 galaxies is not dramatically different from the present one (only a factor $\sim$ 2 lower on average, see e.g. Liang et al. 2004). Hence we will assume that the local correlations remain valid here, at all luminosities, although we note that they most probably break somewhere at higher redshift.
%__________________________________________________________________
   \begin{figure}
   \centering
   \includegraphics[width=9cm]{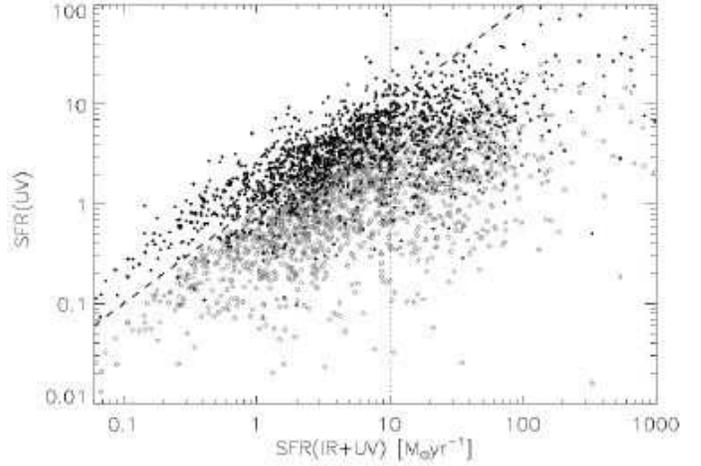}
      \caption{Comparison of the SFR derived from the combined UV+IR light with the SFR derived from either the direct UV light (open circles) or UV corrected for extinction from the $\beta$-slope technique (filled circles).
              }
         \label{FIG:SFRiruv}
   \end{figure}
%__________________________________________________________________

\subsubsection{Computation of the SFR from reprocessed UV, SFR$_{\rm IR}$}
In order to convert the 24\,$\mu$m luminosity into a total IR luminosity (8-1000\,$\mu$m), we used the technique described in Chary \& Elbaz (2001) which is based on the observed correlation between the mid and far infrared luminosity of local galaxies. A library of 105 template spectral energy distribution (SEDs) was built to reproduce those correlations including the tight correlation between the rest-frame 12\,$\mu$m luminosity with L$_{\rm IR}$, probed by the observed 24\,$\mu$m passband for galaxies at $z\sim$1. For each GOODS galaxy, we computed the rest-frame luminosity at 24\,$\mu$m/(1+z) and compared it to the luminosity at that wavelength for each one of the 105 template SEDs to identify the template with the closest luminosity. We then used this template, normalized to the observed luminosity of the galaxy at 24\,$\mu$m/(1+z), to derive L$_{\rm IR}$ for that specific galaxy. This technique was proven to be robust by comparison with the radio continuum at 21 cm, a tracer of massive star formation which correlates tightly with the total IR luminosity of galaxies. The agreement with the radio remains comparable to the error bars on the derivation of L$_{\rm IR}$ from the luminosity at either 21cm or in the mid IR (40\,\%, 1$-\sigma$) up to $z\sim$1 (Elbaz et al. 2002, Appleton et al. 2005). This suggests that the main assumption, i.e. that the local SEDs of galaxies can be applyied to galaxies up to $z\sim$1, is indeed valid. We have assumed here a Salpeter IMF (Salpeter 1955) to keep the same definition of a LIRG or ULIRG as in most studies of these galaxies. Since the UV reprocessed luminosity is a tracer of stars more massive than 5 M$_{\sun}$ typically, a different IMF would lead to different SFR derivations. In the case of a Kroupa IMF (Kroupa 2001), for example, we find that the SFR would be multiplied by 0.7, i.e. SFR(Kroupa)=0.7$\times$SFR(Salpeter IMF). The derived stellar mass is also a function of the IMF and varies with the same factor : M$_{\star}$(Kroupa)=0.7$\times$M$_{\star}$(Salpeter).  As a result, if we adopt a Kroupa IMF instead of a Salpeter one, the specific SFR (SSFR=SFR/M$_{\star}$) would remain unchanged.
Once L$_{\rm IR}$ is determined, we converted it into a SFR$_{\rm IR}$ using the Kennicutt (1998) relation : SFR$_{\rm IR}$ [M$_{\sun}$ yr$^{-1}$] = 1 .72x10$^{-10}$ L$_{\rm IR}$ [L$_{\sun}$]. At a redshift of $z\sim$1, the 5-$\sigma$ point source sensitivity limit of 25\,$\mu$Jy in both GOODS fields can the be translated into a SFR sensitivity limit of 3.5\,M$_{\sun}$ yr$^{-1}$ and sources can be detected down to SFR=1.7 M$_{\sun}$ yr$^{-1}$ (3-$\sigma$ limit).
%__________________________________________________________________
   \begin{figure*}
   \centering
   \includegraphics[width=16cm]{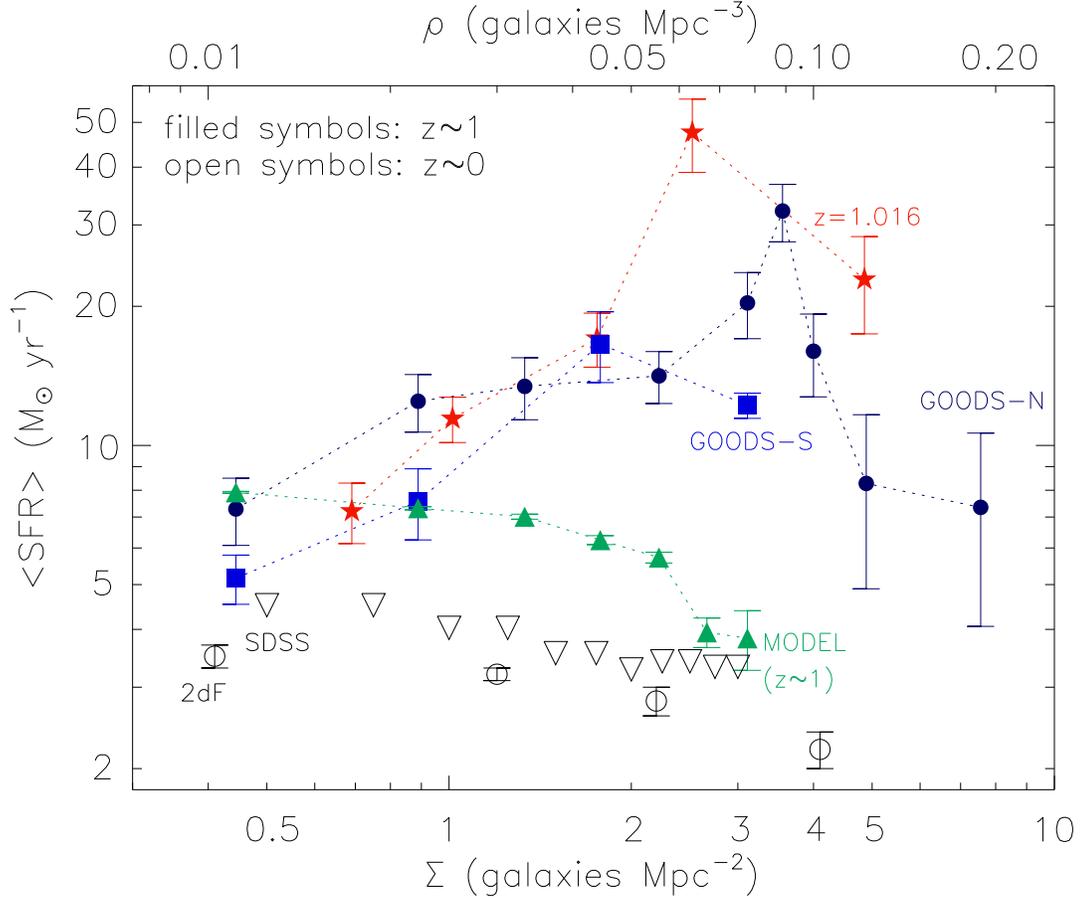}
      \caption{Environmental dependence of SFR at $z\sim$1 and 0. The typical SFR of galaxies is plotted as a function of projected galaxy density, $\Sigma$, computed in boxes of 1.5 comoving Mpc in the Ra,Dec directions and 40 Mpc in the redshift direction ($\Delta$v=$\pm$1500 km s$^{-1}$), see text. The SFR is the sum of SFR$_{\rm IR}$, derived from the 24\,$\mu$m luminosity, and SFR$_{\rm UV}$ (no dust correction). We excluded X-ray identified AGNs from the SFR determination but not $\Sigma$. The filled symbols (dark blue dots: GOODS-N; small blue squares: GOODS-S; red stars: z=1.016 structure in GOODS-N; large green squares: semi-analytical model, Millennium) are for galaxies at z=0.8--1.2, while empty symbols are for the local universe (triangles, SDSS; circles, 2dF). The local and Millennium samples are limited to M$_{\rm B}$=-20 to mimick the selection at z$_{\rm AB}$=23.5 in GOODS. SFR in the SDSS are derived from H$_{\alpha}$ luminosity corrected for extinction (Brinchmann et al. 2004) and by H$_{\alpha}$ equivalent widths in the 2dF (Lewis et al. 2002; SFR normalized to L$_{\star}$, see their Fig.8).  Vertical error bars were derived from bootstrapping. We used a Salpeter stellar initial mass function (IMF) to compute the SFR and stellar masses. 
              }
         \label{FIG:mainfig}
   \end{figure*}
%__________________________________________________________________

\subsubsection{Computation of the complete SFR census from reprocessed and direct UV}
In order to obtain a complete census of the SFR in galaxy populations, we have also computed the SFR$_{\rm UV}$ associated to the direct UV light radiated by the galaxies and not reprocessed by dust. For this purpose, we used the observed B-band that, at a redshift of $z\sim$1, corresponds to the near UV in the rest-frame at 2200\,\AA. The conversion of L$_{\rm UV}$ to SFR$_{\rm UV}$ was done using the Eq.5 of (Daddi et al. 2004) for a Salpeter IMF (Eq.~\ref{EQ:SFRUV} here).
\begin{equation}
\label{EQ:SFRUV}
{\rm SFR}_{\rm UV}~[{\rm M}_{\sun}~{\rm yr}^{-1}] = L_{\rm UV} [{\rm erg~s^{-1}~Hz^{-1}}]~/~(8.85 \times 10^{27})
\end{equation}
The depth of the B-band data from HST+ACS  allows us to extend the census of star-formation activity down to SFR$_{\rm UV}$ limits of ~0.5 M$_{\sun}$ yr$^{-1}$.

We explicity demonstrate here how the direct measurement of the IR emission of galaxies (in the dust regime) is a crucial ingredient to study the SFR of distant galaxies. When compared to our estimate of the SFR(IR+UV) (=SFR$_{\rm IR}$+SFR$_{\rm UV}$), it is found that the SFR$_{\rm UV}^{\rm corr}$ derived from the UV corrected for extinction luminosity using the technique of the $\beta$-slope (Meurer, Heckman \& Calzetti 1999; where F$_{\lambda}\sim \lambda^{\beta}$) saturates around 10 M$_{\sun}$yr$^{-1}$, while the SFR(IR+UV) varies by three orders of magnitudes up to $\sim$100 M$_{\sun}$ yr$^{-1}$ (Fig.~\ref{FIG:SFRiruv}). Here we used the observed U-B color to compute $\beta$ and the apparent B-band magnitude to compute L$_{\rm UV}$.

The $\beta$-slope technique relies on the fact that, without dust extinction, the composite UV spectrum of a galaxy is flat over the wavelength range 1500-3000\,\AA\,range in F$_{\nu}$  (see e.g. Kennicutt 1998), hence $\beta$=-2 in the case of no dust extinction. The $\beta$-slope was found to correlate with the FIR over UV flux density ratio of a sample of starburst galaxies, suggesting that knowing L$_{\rm UV}$ and $\beta$ would permit to derive L$_{\rm IR}$. However, after the launch of the GALEX satellite, the sample of galaxies with measured L$_{\rm UV}$ and $\beta$ has largely grown and it was shown that the correlation does not work both for normal galaxies and for IR selected galaxies (Buat et al. 2005, see also Goldader et al. 2002). This can probably be understood by a geometry effect: the IR light comes from the dense and dust obscured giant molecular clouds and associated photo-dissociation regions while the UV light extends over a wider area in galaxies, including regions with less extinction. 

%__________________________________________________________________
   \begin{figure}
   \centering
   \includegraphics[width=9cm]{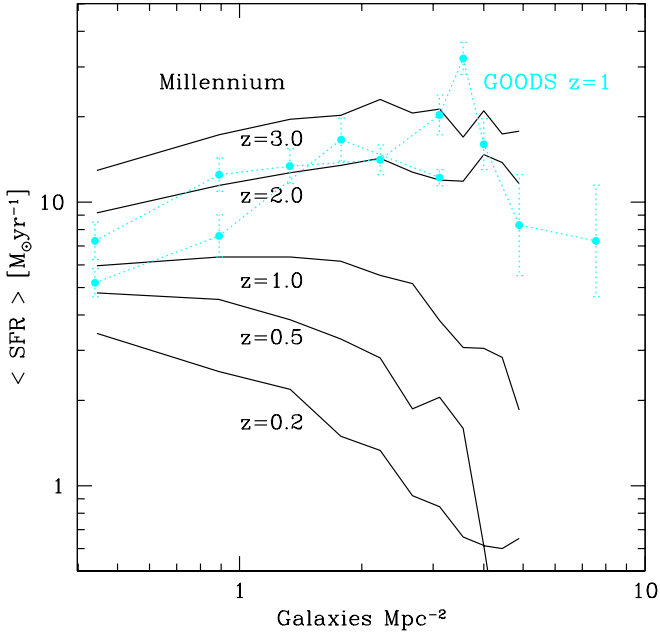}
      \caption{Environmental dependence of SFR in the Millennium model in five redshift slices of the Universe from z=0.2 to 3. The trends observed in the two GOODS fields at z=1 are shown with blue dotted lines for comparison. We applied the same selection criterion that we used for GOODS (M$_{\rm B}\leq$ -20) to galaxies in the simulations. We computed $\Sigma$, the local galaxy density, with the same technique as for GOODS. 
              }
         \label{FIG:mille_z}
   \end{figure}
%__________________________________________________________________~\ref{FIG:mainfig}
%__________________________________________________________________
   \begin{figure}
   \centering
   \includegraphics[width=8cm]{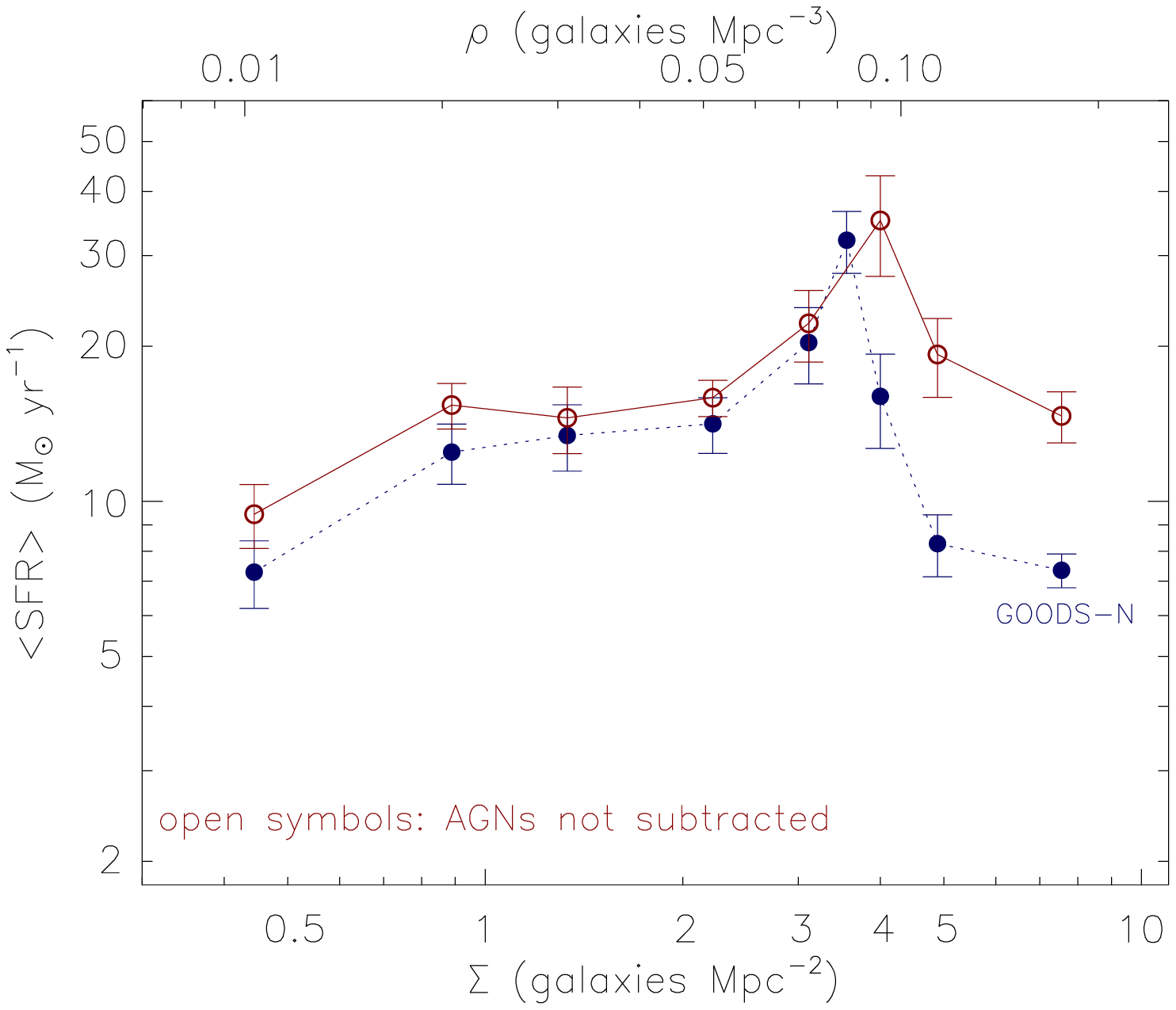}
    \includegraphics[width=8cm]{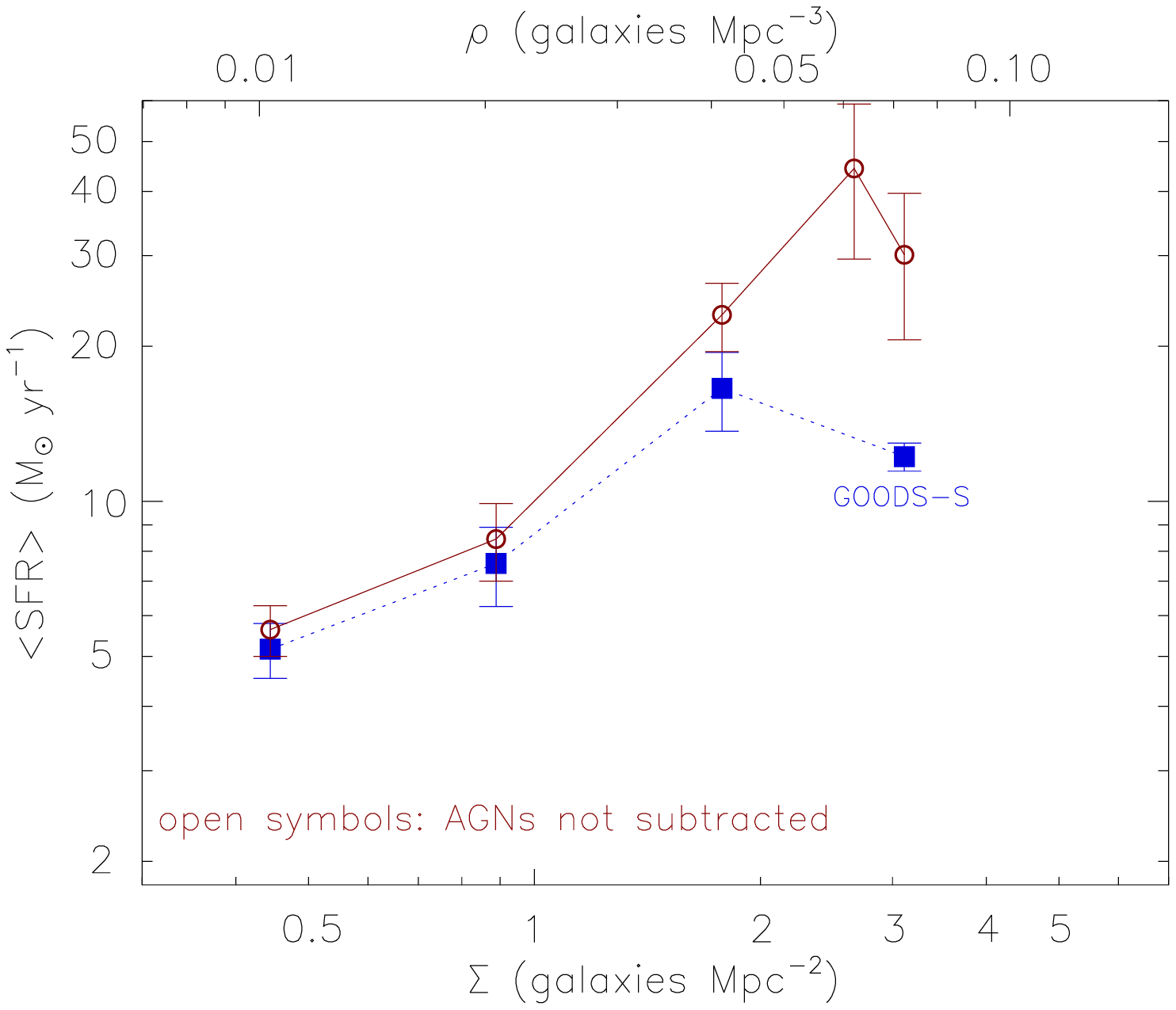}
   \caption{Effect of the AGN contribution on the SFR-density relation in GOODS-N(top; 14\,\% of AGNs within 0.8$\leq z \leq$ 1.2 and z$_{\rm AB}\leq$ 23.5) and GOODS-S (bottom; 12\,\% of AGNs within 0.8$\leq z \leq$ 1.2 and z$_{\rm AB}\leq$ 23.5). The filled symbols mark the relation used to produce Fig.~\ref{FIG:mainfig} for which we used all galaxies in the computation of $\Sigma$, the projected galaxy density, but where we set the SFR of galaxies with an AGN to zero. The open symbols represent the other extreme, where we assume that the light from those galaxies is due to star formation only.
              }
         \label{FIG:AGN_GN}
   \end{figure}
%__________________________________________________________________

%__________________________________________________________________
   \begin{figure*}
   \centering
   \includegraphics[width=16cm]{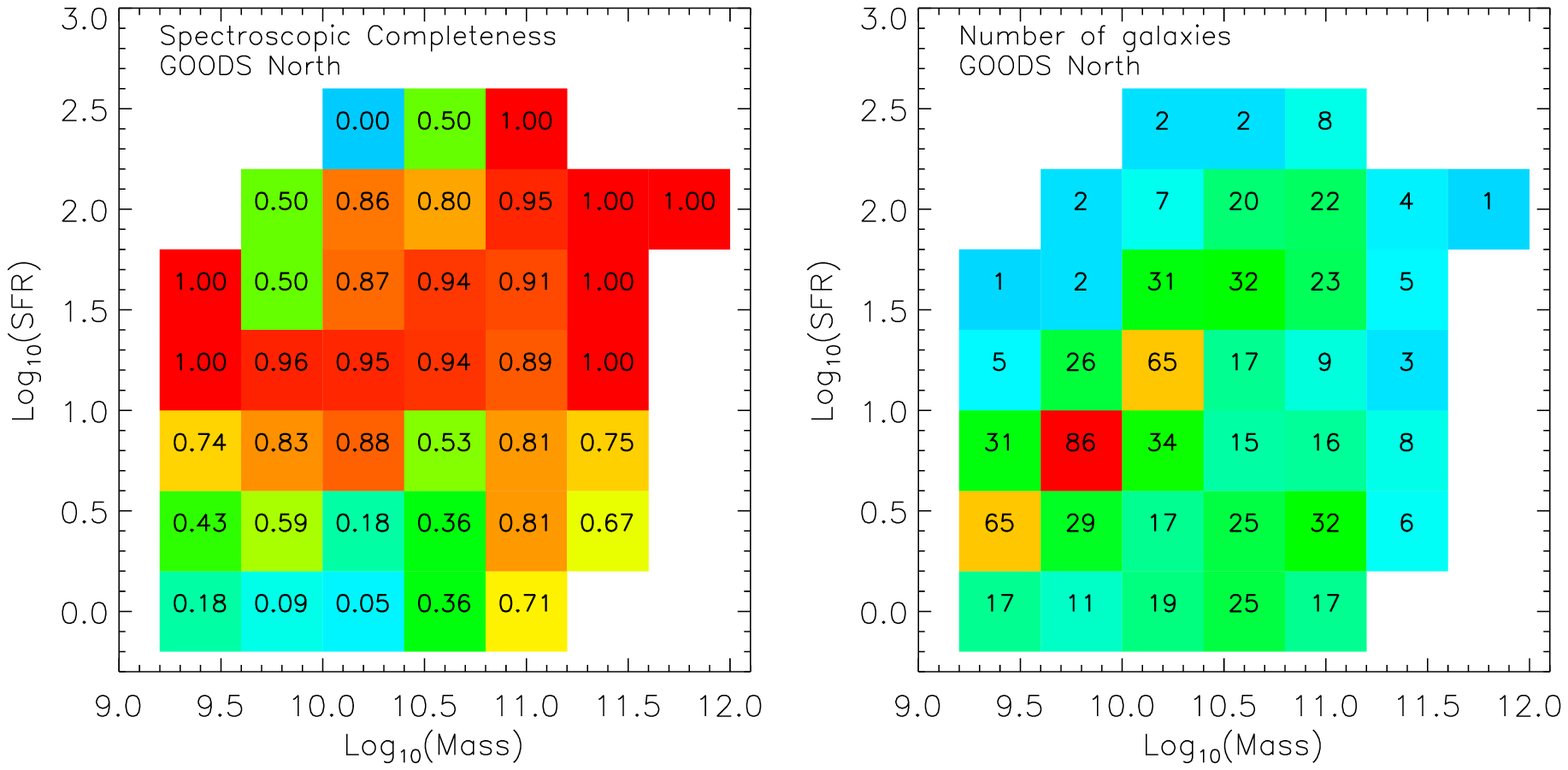}
      \caption{(left) Spectroscopic completeness as a function of SFR and stellar mass for galaxies with z$_{\rm AB}\leq$23.5 and 0.8$\leq z_{\rm phot} \leq$1.2 in GOODS-N. (right) Number of galaxies within each (SFR, M$_{\star}$) cell and with [z$_{\rm AB}\leq$23.5, 0.8$\leq z_{\rm phot} \leq$1.2].
              }
         \label{FIG:SFRcompGN}
   \end{figure*}
%__________________________________________________________________
%__________________________________________________________________
   \begin{figure*}
   \centering
   \includegraphics[width=16cm]{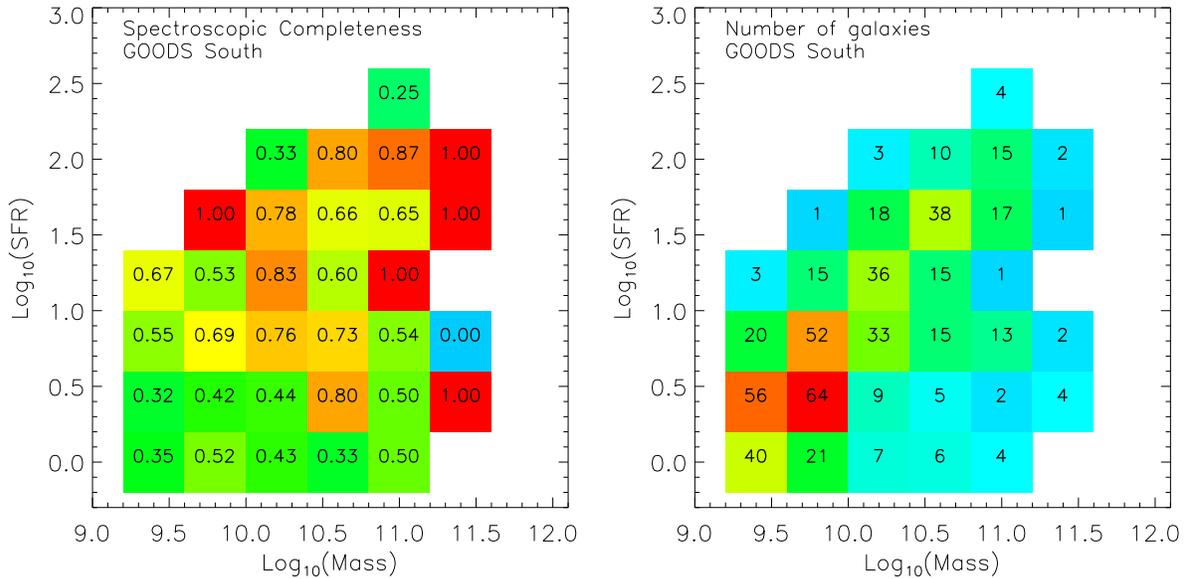}
      \caption{(left) Spectroscopic completeness as a function of SFR and stellar mass for galaxies with z$_{\rm AB}\leq$23.5 and 0.8$\leq z_{\rm phot} \leq$1.2 in GOODS-S. (right) Number of galaxies within each (SFR, M$_{\star}$) cell and with [z$_{\rm AB}\leq$23.5, 0.8$\leq z_{\rm phot} \leq$1.2].              }
         \label{FIG:SFRcompGS}
   \end{figure*}
%__________________________________________________________________

\section{The reversal of the star formation--density at $z\sim$1, comparison to the Millennium and SDSS}
\label{SEC:result}
We show here that the familiar, local SFR-density relation is reversed at $z\sim$1, with much larger typical SFR per galaxy in dense environments with respect to low density environments (Fig.~\ref{FIG:mainfig}). In both fields, the average SFR of galaxies increases by factors of 3-4, when the projected galaxy density increases from $\Sigma$=0.4 to $\sim$3 gal Mpc$^{-2}$. The probability that the observed increase is obtained by chance is estimated to be lower than 10$^{-4}$ for both fields using Monte Carlo realizations. We also checked that this trend is robust against selection effects, such as differential spectroscopic incompleteness (see Sect.~\ref{SEC:completeness}). Correcting for incompleteness (which more strongly affects galaxies with low SFR and stellar mass), or selecting only blue-star forming galaxies, would in fact enhance the rise of $<$SFR$>$ with density. The decrease of the $<$SFR$>$, observed beyond its peak at $\sim$3-4 gal Mpc$^{-2}$, is consistent with the properties of galaxies in distant, rich clusters (Rosati et al. 1999) and suggests a ÒdownscalingÓ scenario, where the typical environmental density at which the $<$SFR$>$ peaks decreases with the passing of cosmic time.

The SFR-density trend observed at $z\sim$1 is opposite to the local one that we have estimated using the same technique applied to the SDSS, where the SFR was derived from the H$_{\alpha}$ luminosity corrected for extinction (Binchmann et al. 2004), after applying the same selection criterion of M$_{\rm B}\leq$-20 (Fig.~\ref{FIG:mainfig}). This local relation that we derived from the SDSS is consistent with the one observed in the 2 degree Field (2dF) survey (again with M$_{\rm B}\leq$-20; see Lewis et al. 2002). When compared to GOODS at $z\sim$1, the SFR in both local surveys is less than twice smaller at low densities but changes by an order of magnitude in dense environments.

We applied the same technique and selection criterion (M$_{\rm B}\leq$ -20) to a simulated lightcone of the $z\sim$1 universe, within the framework of the $\Lambda$CDM hierarchical clustering models (Millennium). We find that the average comoving density of galaxies inside 0.8$\leq z \leq$1.2 in the Millennium model matches closely that in the GOODS fields. However, the predicted SFR--density relation is opposite to that observed in GOODS, with the $<$SFR$>$ declining with increasing density. The model does predict an increase of the SFR with density but only at $z \geq$2, where it is much weaker than the one observed at $z\sim$1 (Fig.~\ref{FIG:mille_z}).  

We have also used the Millennium lightcones to investigate the robustness of our result against bias due to larger spectroscopic incompleteness for galaxies with low SFRs.  As an extreme approach, we excluded from the analysis all simulated galaxies forming less than 3 M$_{\sun}$yr$^{-1}$ (disfavored by the spectroscopy). Even there, we still find a predicted SFR--density trend that is substantially flat, similar to the one shown in Fig.~\ref{FIG:mille_z}. We did the same test on the SDSS galaxy sample and found a similar result. As a further test, we extracted 150 independent GOODS-sized fields at 0.8$\leq z \leq$1.2 from the Millennium to check whether the reversal we detect could be actually due to cosmic variance combined with the limited size of the survey. We never found an increase of SFR with density comparable with the ones seen in the two GOODS fields.

\section{Discussion of the systematics in the determination of the average SFR and local galaxy density}
\label{SEC:systematics}
\subsection{The role of AGNs}
\label{SEC:AGNs}
One of the main strengths of GOODS is that these two fields are located inside the X-ray Chandra deep fields (2 Msec in GOODS-N, 1 Msec in GOODS-S centered on the Chandra Deep Field North \& South respectively). Active galactic nuclei (AGNs) were identified in both fields by Bauer et al. (2004) as the galaxies with either N$_{\rm H}\geq$10$^{22}$ cm$^{-2}$, a hardness ratio greater than 0.8 (ratio of the counts in the 2-8 keV to 0.5-2 keV passbands), L$_{\rm X}$[0.5-8.0 keV] $>$ 3$\times$10$^{42}$ ergs s$^{-1}$, or broad / high-ionization AGN emission lines. As mentioned in Bauer et al. (2004), some AGNs might be missed by this technique, however such AGNs are difficult to identify even locally and we have adopted here a conservative approach (described below) which would counterbalance this effect.
Since there is no robust technique to separate the contribution from star formation and the AGN to the mid-infrared emission of a galaxy, especially for distant galaxies for which we only have a few broadband magnitudes and no spectrum in the infrared, we  conservatively assume that the 24\,$\mu$m emission from any galaxy harbouring an AGN is predominantly powered by the active nucleus, and its SFR is set to zero. 
In the redshift range 0.8$\leq z \leq$1.2, the fraction of AGNs is 9 and 12\,\% in GOODS-N and GOODS-S respectively (among galaxies with a spectroscopic redshift and z$_{\rm AB}\leq$ 23.5, i.e. M$_{\rm B}\leq$-20). Assuming no star formation at all in AGNs places a lower-limit on their role in the SFR-environment relation.  We have compared this to the other extreme case, where we assume that the mid-infrared emission is entirely due to star formation, with no contribution from the AGN (see Figure 8). The trend remains similar to the one presented in the paper in both cases. The true behaviour is most probably locate inside the enveloppe defined by these two extreme scenarios. Finally, we note that all galaxies were used in the computation of $\Sigma$, independantly of the AGN contribution.

\subsection{Selection effects due to the spectroscopic incompleteness}
\label{SEC:completeness}
If we separate galaxies in the reference sample (i.e. with a photometric redshift within 0.8$\leq z \leq$1.2) as a function of their SFR and stellar mass, we find that the galaxies suffering most from the spectroscopic incompleteness are those with small SFR and stellar masses (see Figs.~\ref{FIG:SFRcompGN},\ref{FIG:SFRcompGS}). Galaxies with a stellar mass (M$_{\star}$) larger than 5$\times$10$^{10}$ M$_{\sun}$ exhibit a completeness of $\sim$80\,\% while low SFR and M$_{\star}$ galaxies are $\sim$50\,\% complete and even less in the lowest SFR bin of galaxies with nearly no star formation. In order to test the robustness of the increase of the SFR with projected galaxy density presented in the Fig.~\ref{FIG:mainfig}, we have mimicked the lower level of incompleteness in the lower-left corner of Figs.~\ref{FIG:SFRcompGN},\ref{FIG:SFRcompGS} by undersampling randomly the galaxies located in the most homogeneous region (i.e. SFR$\geq$10 M$_{\sun}$yr$^{-1}$ and M$_{\star} \geq 5 \times 10^{10}$ M$_{\sun}$). The net effect of this process is that it increases the error bars but keeps the same increase of the SFR versus projected galaxy density. Alternatively, we found that the trend also remained present (with larger error bars) by selecting only galaxies with M$_{\star}\geq 5\times 10^{10}$ M$_{\sun}$, where the completeness is the highest. 

The trend found in Fig.~\ref{FIG:mainfig} is therefore not produced by a selection effect associated with the spectroscopic incompleteness. More precisely, the observed increase of star formation with galaxy density at $z\sim$1 is most probably underestimated, since low-mass galaxies with little to no star formation activity lie in the regions with the lowest galaxy density (see Fig.~\ref{FIG:Mstardens}). Hence if we were to correct for incompleteness, we would need to lower the average SFR in the lowest density bins, which would then result in a greater increase of the average SFR from low to high galaxy density bins. We conclude that the trend presented in Fig.~\ref{FIG:mainfig} is real and not artificially produced by a selection effect due to incompleteness in the spectroscopic redshift sample.

\section{Comparison to the evolution of the color bimodality of galaxies}
\label{SEC:bimodality}
%__________________________________________________________________
   \begin{figure}
   \centering
   \includegraphics[width=8cm]{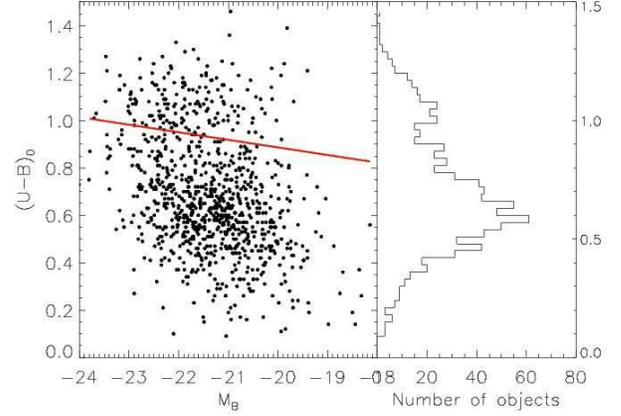}
      \caption{Rest-frame colour-magnitude diagram for the GOODS galaxies with $z$=0.8--1.2. The plain line separating red and blue galaxies follows the definition used for the DEEP2 survey (Cooper et al. 2007).
              }
         \label{FIG:GOODScolmag}
   \end{figure}
%__________________________________________________________________
%__________________________________________________________________
   \begin{figure}
   \centering
   \includegraphics[width=8cm]{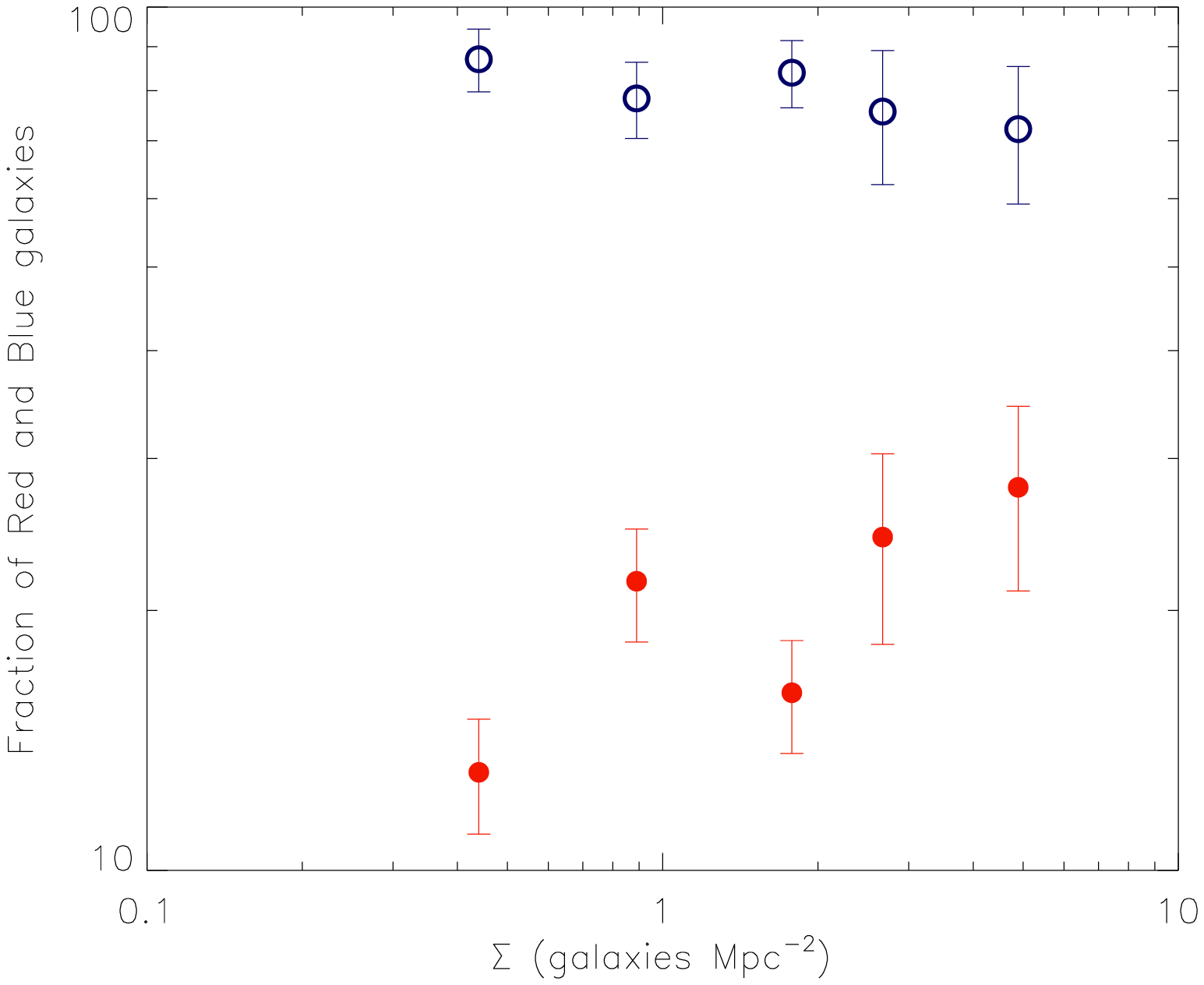}
   \includegraphics[width=8cm]{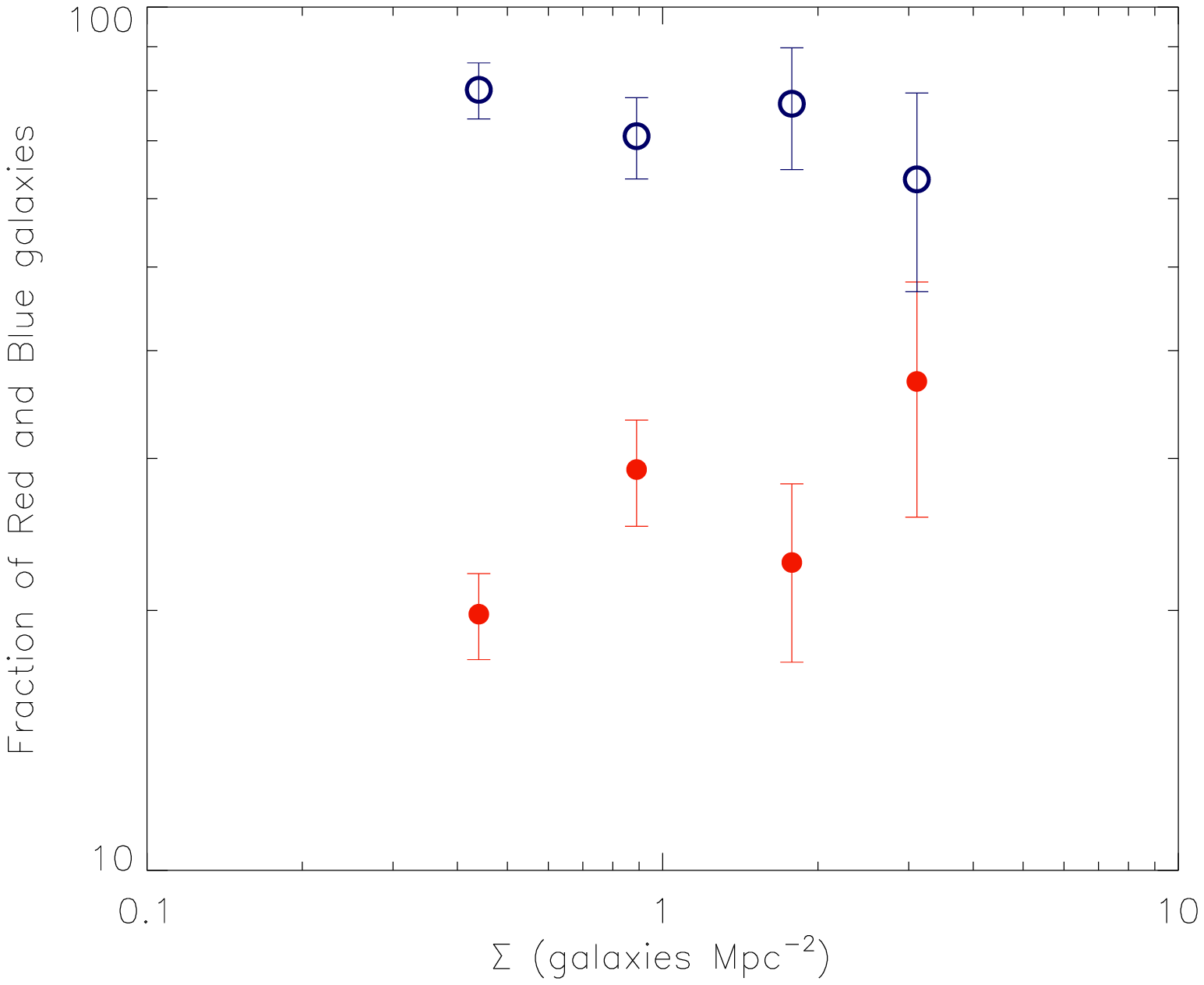}
      \caption{Fraction of red and blue galaxies in GOODS-N (top) and GOODS-S (bottom) as a function of projected galaxy density at $z$=0.8--1.2. The total fraction of red galaxies is 17\,\% (GOODS-N) and 24\,\% (GOODS-S) respectively for galaxies with z$_{\rm AB}\leq$23.5.
                    }
         \label{FIG:bimod}
   \end{figure}
%__________________________________________________________________
%__________________________________________________________________
   \begin{figure}
   \centering
   \includegraphics[width=8cm]{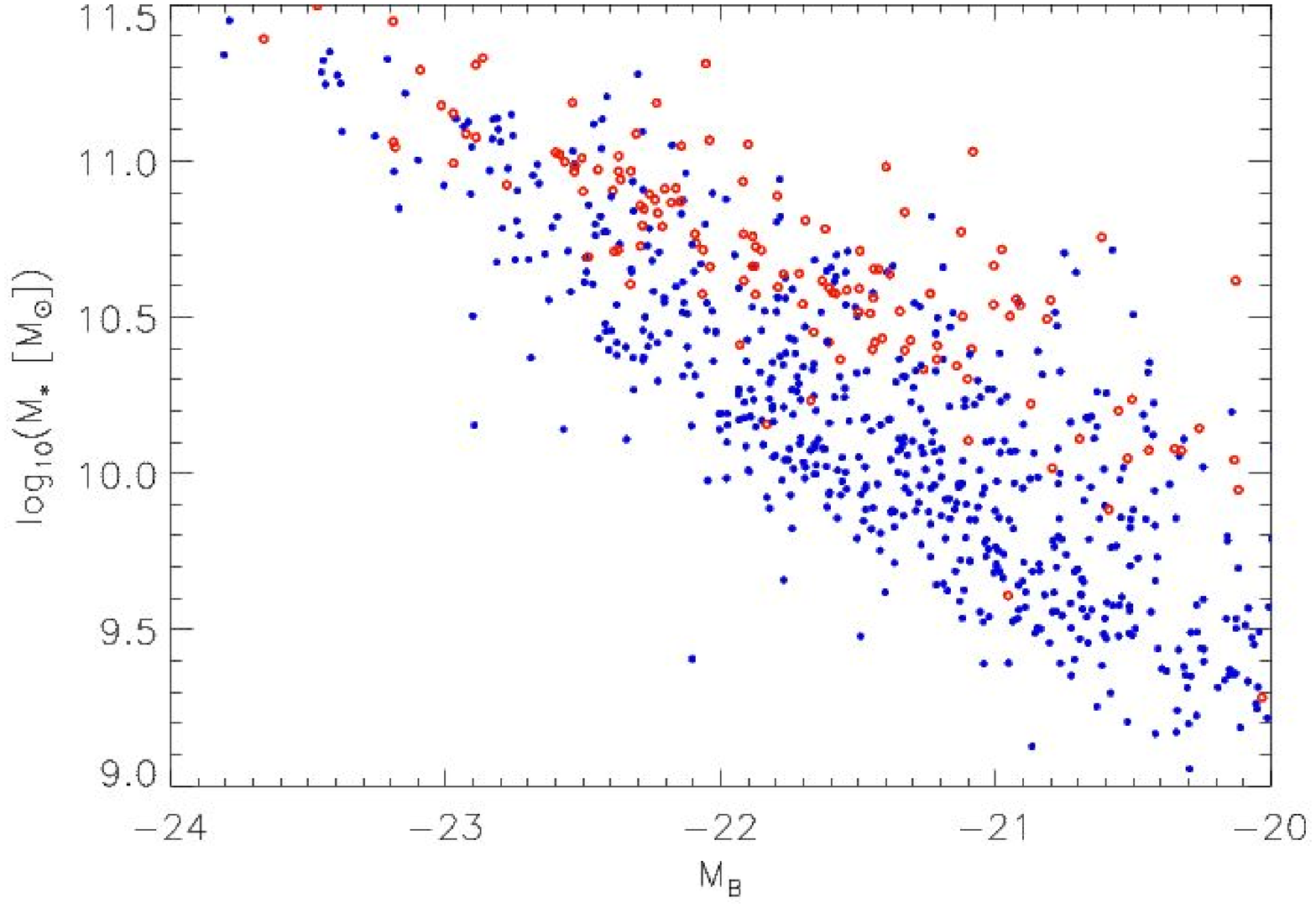}
    \includegraphics[width=8cm]{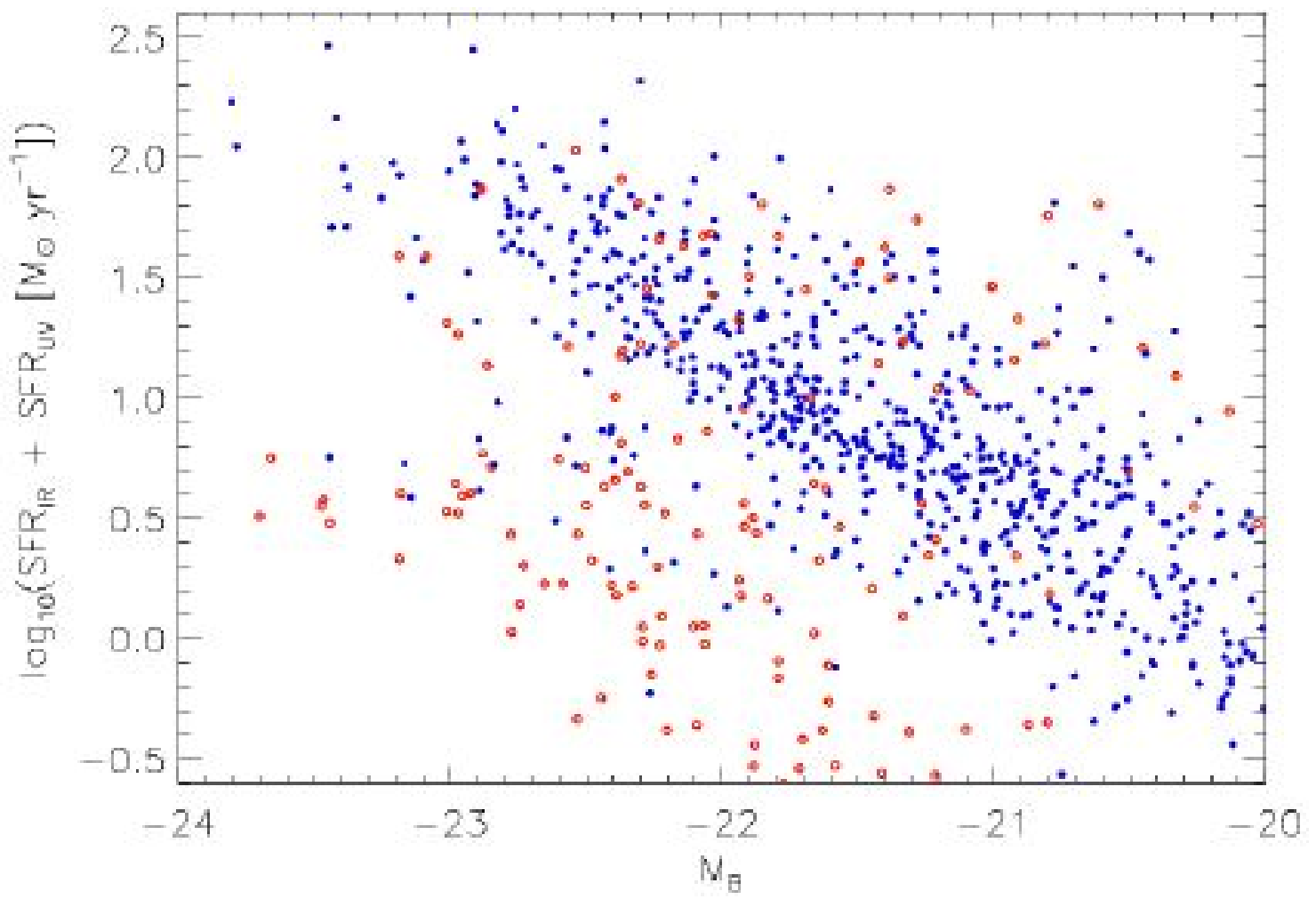}
     \caption{Absolute B-band magnitude versus stellar mass (top) and star formation rate (SFR= SFR$_{\rm IR}$+SFR$_{\rm UV}$; lower plot) of GOODS-N and GOODSÐS galaxies with $z$=0.8--1.2. 
                    }
         \label{FIG:Mstar_Mb}
   \end{figure}
%__________________________________________________________________
Recent studies have addressed the star formation-density relation from the angle of the color bimodality of galaxies, by following the evolution of the fraction of red galaxies with galaxy density (Bell et al. 2004, Cooper et al. 2006, Cucciati et al. 2006). At first glance, our result might appear to be in contradiction with their results, since it appears that at $z\sim$1, like at $z\sim$0, high galaxy densities are the preferred regions of red-dead galaxies, by opposition to blue-actively star forming galaxies. Here, we show that the GOODS galaxies used to produce the Fig.~\ref{FIG:mainfig} do also follow a bimodal color distribution in the rest-frame U-B versus M$_{\rm B}$ color-magnitude diagram (see Fig~\ref{FIG:GOODScolmag}), that is marked by the same formula as the one used by Cooper et al. (2007) for the DEEP2 Galaxy Redshift Survey (Eq.~\ref{EQ:limitBimod}), where rest-frame U and B magnitudes were derived from the fit of their observed spectral energy distributions. 
\begin{equation}
\label{EQ:limitBimod}
U-B = -0.032(M_{\rm B}+21.52)+0.454-0.25+0.831
\end{equation}
We also confirm with the same sample of galaxies that we used to produce the Fig.~\ref{FIG:mainfig} that the fraction of red galaxies increases with galaxy density in the redshift interval 0.8$\leq z \leq$1.2 (Fig.~\ref{FIG:bimod}). However, we note that at $z\sim$1, the fraction of red galaxies is only 17 and 24\,\% in GOODS-N and GOODS-S respectively, in perfect agreement with the DEEP2 survey (see Cooper et al. 2007, Gerke et al. 2006). Therefore the increase of the average SFR of galaxies with galaxy density that we present here at $z\sim$1 is not in contradiction with the increasing fraction of red galaxies with density: first because this fraction remains low at all galaxy densities at these epochs, second because some fraction of the distant red population might be experiencing dust obscured star formation. It is also consistent with the finding that the blue luminosity of the most luminous blue galaxies increases with galaxy density (Cooper et al. 2006), which can now be understood as probably partly due to their increased star formation activity. However, the blue luminosity alone is not sufficient to solve this issue since it is a better tracer of stellar mass than star formation (see Fig.~\ref{FIG:Mstar_Mb}) and since the stellar mass of galaxies does also increase with galaxy density (see Fig.~\ref{FIG:Mstardens}).

It is important to note that when considered alone, the progressive decline of the fraction of blue galaxies can be understood as evidence that star formation is progressively quenched in massive galaxies which pass from the blue side to the red side of the road of galaxy evolution. But we find here that the regions that contain today the largest 
fraction of dead galaxies, were the ones which harbored most star formation in the past. This suggests that the mechanism which relates galaxies to their environment is not simply a quencher but also a trigger of star formation. Both apparently contradictory mechanisms would be naturally explained if star formation was accelerated by the environment of galaxies, resulting in the faster exhaustion of their gas reservoir. 
%__________________________________________________________________
   \begin{figure*}
   \centering
   \includegraphics[width=11cm]{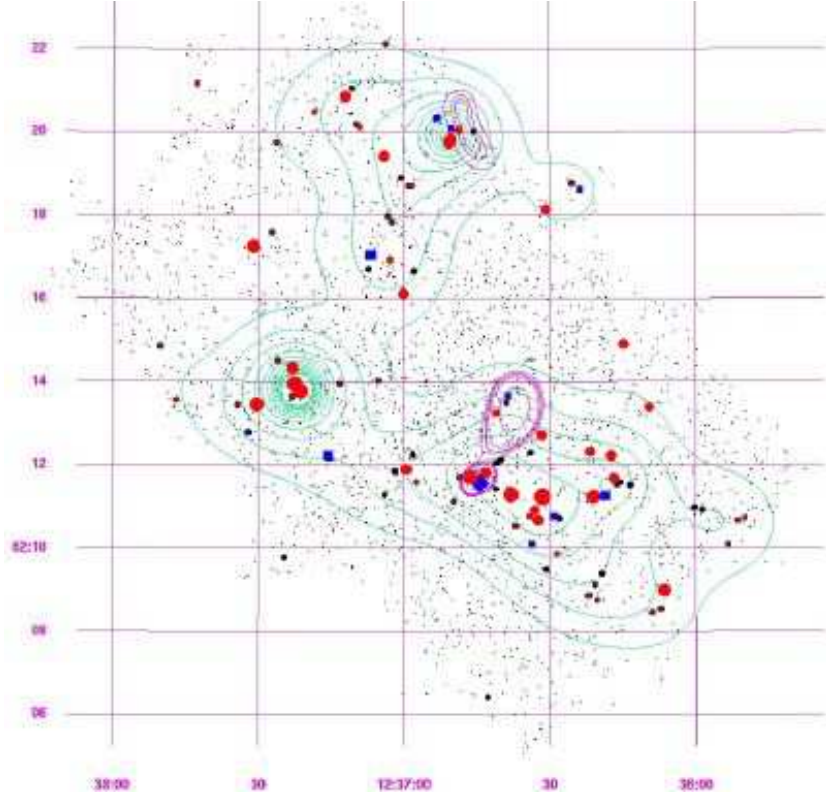}
      \caption{Spatial distribution of the GOODS-N galaxies (large symbols) in the z=1.0163 [1.0074, 1.0254] structure (10$\arcmin \times$ 16$\arcmin$, i.e. 10$\times$16 comoving Mpc). Red: MIPS-24\,$\mu$m detections (red filled circles proportional to the total IR luminosity, i.e. SFR). Blue squares (size proportional to total IR luminosity): X-ray identified AGNs (only 3 are undetected at 24\,$\mu$m). Black filled circles: remaining galaxies at z=1.0163 undetected in IR and not X-ray identified as AGNs. Small dots: whole sample of 2162 GOODSN galaxies with a spectroscopic redshift. Blue diamond: FRI galaxy (VLA--J123644.3$+$621133) at z=1.0128. Green contours: isodensity profiles from the wavelet filtering of the image produced by the distribution of the 107 galaxies located at $z\sim$1.0163. The 12 contours correspond to densities ranging from 0.4 to 4.9 galaxies Mpc$^{-2}$ (step= 0.4 gal.Mpc$^{-2}$). Purple contours: contours of the X-ray extended sources coinciding with 2 groups of galaxies. 
                    }
         \label{FIG:contours}
   \end{figure*}
%__________________________________________________________________

\section{Discussion}
\label{SEC:discussion}
In this section, we discuss the origin of the reversal of the star formation-density relation at $z\sim$1 found in Fig.~\ref{FIG:mainfig}. A major difficulty that is encountered when trying to understand the role of the environment on galaxy activity is that denser environments contain preferentially more massive galaxies. It is therefore difficult to separate the effect of external conditions to the increased in-situ activity due to a larger galaxy mass. Yet, this might not be a real issue in the framework of hierarchical galaxy formation. A massive galaxy being born from the successive fusion of many sub-entities, what is now internal to a galaxy was its external close environment in the past. This leads us to make a distinction between the local environment of a galaxy and large-scale structures, which in terms of the hierarchical scenario would translate into different dark matter halo masses, that of a galaxy and that of a group or a cluster.

In Sect.~\ref{SEC:cluster}, we present the case of a candidate proto-Virgo cluster at $z=$1.016 to study the connection of the reversal of the star formation-density relation at $z\sim$1 with large-scale structure formation. In Sect.~\ref{SEC:mergers}, we study the role of major mergers in the triggering of LIRGs in this structure from their HST-ACS morphology.

In Sect.\ref{SEC:SFRMstar}, we present a correlation between the SFR and stellar mass of galaxies, which could result in the trend observed in Fig.~\ref{FIG:mainfig}, if the stellar mass of galaxies was increasing enough with galaxy density.
Finally, in Sect.\ref{SEC:SSFR}, we find evidence that this is probably not the case by studying the dependence of the specific SFR (SSFR=SFR/M$_{\star}$) with galaxy density.

\subsection{Role of structure formation : the case study a candidate proto-Virgo cluster at $z=$1.016}
\label{SEC:cluster}
The possible connection between the SFR of galaxies and structure formation can be directly tested in one of the major redshift peaks of GOODS-N at $z=$1.016$\pm$0.009. The contours of the wavelet decomposition of the projected position of the 107 galaxies with a spectroscopic redshift belonging to this structure are presented in Fig.~\ref{FIG:contours}.

We have quantified the non-uniformity of the galaxy distribution at $z\sim$1.016 by comparing the real galaxy positions to 1000 Poissonian realizations with 107 galaxies extracted from the total sample of galaxies with a spectroscopic redshift. We applied a wavelet decomposition to both the real and fake galaxy distributions using seven wavelet scales (scale 1 has a 35 comoving-kpc size, while scale 7 has a 35 kpc $\times$ 26 = 2.2 comoving-Mpc size),  and computed the kurtosis of each scale.  We subtracted the mean kurtosis over the 1000 Monte-Carlo realizations to the one measured on each wavelet scale of the real image and divided the result by the rms dispersion of the MC realizations to quantify the non uniformity at each scale. While below a 500 kpc wavelet size, the signal-to-noise ratio is lower than 3, it rises to S/N=14 at the 0.3 Mpc size and decreases below (S/N=8 at 0.56 Mpc, 2 at 1.1 Mpc and 7 at 2.2 Mpc), clearly suggesting an excess of structures at the scale typical of galaxy groups. 
%__________________________________________________________________
   \begin{figure}
   \centering
   \includegraphics[width=9cm]{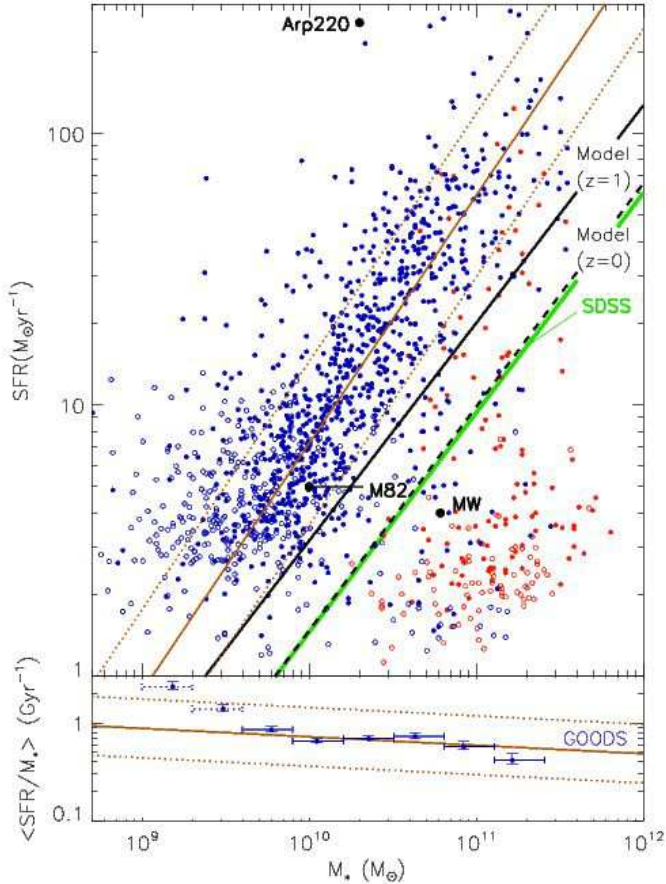}
      \caption{Relationship between SFR and specific SFR (SFR/M$_{\star}$) with stellar mass in $z\sim$0.8-1.2 galaxies. The SFR is the sum of the SFR(UV) and SFR(IR) when galaxies are detected at 24\,$\mu$m (filled symbols) and SFR(UV) only otherwise (open symbols). A Salpeter IMF is assumed. Stellar masses are derived from models fit to ground-based ultraviolet through near-infrared photometry from the KPNO, CTIO and Subaru observatories and Spitzer IRAC imaging at 3.6 and 4.5\,$\mu$m using PEGASE.2 for a Salpeter IMF. Red and blue symbols are for red and blue galaxies as defined in Fig.\ref{FIG:GOODScolmag}. The GOODS blue galaxies follow the relation (plain brown line): SFR[M$_{\sun}$ yr$^{-1}$]=7.2 [-3.6,+7.2]$\times$(M$_{\star}$/10$^{10}$ M$_{\sun}$)$^{0.9}$ at the 68\,\% confidence level (marked with dashed brown lines). 
              }
         \label{FIG:SFRMstar_GOODS}
   \end{figure}
%__________________________________________________________________

We identify three main groups of galaxies at the $\sim$1-2 comoving Mpc scale, two of which are associated with extended X-ray emission (Bauer et al. 2002) indicating that they are gravitationally bound. The probability that these groups result from a random distribution of galaxies is $\sim$0.2\,\% (2 cases among the 1000 MC realizations). The largest group (lower left in Fig.~\ref{FIG:contours}) is centered close to a 0.5 Mpc (comoving, i.e. 30 arcseconds) extended X-ray detection likely to be a group or poor cluster of galaxies (CXOHDFN J123645.0+621142, Bauer et al. 2002), radiating an X-ray luminosity of L$_X$[0.5-Ð2 keV]=2$\times$10$^{42}$ erg s$^{-1}$ (absorption-corrected rest-frame luminosity) as measured from the 1 mega-seconds Chandra X-ray image of this region (CDFN). The existence of such a group in GOODS within 0.8$\leq z \leq$1.2 is perfectly consistent with expectations as summarized in Eq.~\ref{EQ:NbX}. The fact that the X-ray emission results from the thermal brehsstrahlung radiation of the ICM of a group of galaxies is reinforced by the presence of a Fanaroff-Riley type I (FRI) radio galaxy (VLA--J123644.3$+$621133; Richards et al. 1998) at $z=$1.0128, which are typically found in the centers of galaxy clusters. 

This group is asymetric and extended in the direction of the most eastern group on the left of the image as well as of the northern group. This suggests the possibility that the groups belong to a larger structure (see Fig.~\ref{FIG:contours}), similar to the filaments observed in the nearby universe at the intersection of which galaxy clusters are expected to form. We also find evidence for a filamentarity distribution by computing the maximum number of points which can be connected using a continuous function which slope remains below one on a Lipschitz graph. We get a total of 22 connected galaxies with 0.6\,\% chance of being random. 

These results suggest that we are witnessing the birth of a galaxy cluster with a central region becoming virialized and responsible for the extended X-ray emission, while galaxies are continuously flowing through the filament towards the cluster. The total stellar mass of the detected galaxies (2$\times$10$^{12}$ M$_{\sun}$) would then lead to a total mass $\geq$10$^{14}$ M$_{\sun}$ (including dark matter) similar to that of the Virgo cluster (see Fig.6 of Lin, Mohr \& Stanford 2003, for M$_{\star}$/L$_{\rm K}$=1). 

The structure contains 16 AGNs (=15\,\%) and 25 LIRGs (=26\,\%), the remaining objects being galaxies forming stars at rates below $\sim$20 M$_{\sun}$ yr$^{-1}$. Only 4 of the galaxies in the structure are also detected in the radio at 1.4 GHz and the ratio of the total IR luminosity derived from the radio and mid IR for those galaxies is prefectly consistent (=1.1$\pm$0.3). 

The most actively star forming galaxies in this structure, the LIRGs, are nearly all located at the centers of the three main galaxy groups. We also find that their intrinsic luminosity increases as a function of the local density of galaxies. The size of the red dots in the Fig.~\ref{FIG:contours} is proportional to the total IR luminosity of the galaxies. This structure hence offers a direct visualization of the trend obtained in Fig.~\ref{FIG:mainfig}. This strongly suggests that a possible mechanism to explain the observed reversal of the star formation-density relation is that the SFR of galaxies have been accelerated during group formation at the $\sim$1 Mpc scale. We also note that while galaxy clusters contain only 5 to 10\,\% of present-day galaxies, most local galaxies belong to groups such as those detected here (although generally less clustered together). The physical mechanism through which groups could affect galaxies remains to be understood. In the next section we discuss the specific case of major mergers.

%__________________________________________________________________
   \begin{figure}
   \centering
   \includegraphics[width=8cm]{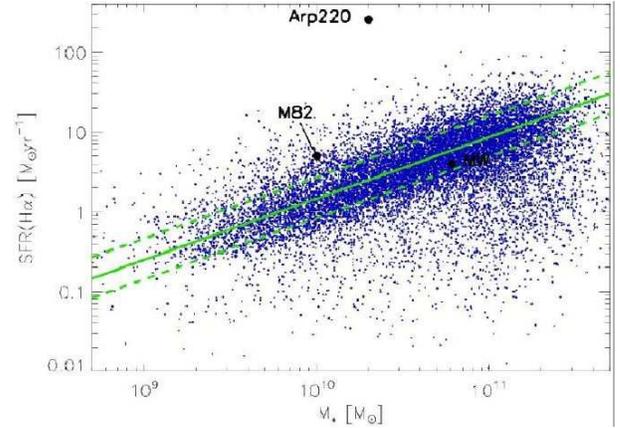}
      \caption{Correlation between the SFR derived from H$_{\alpha}$ corrected for extinction and the stellar mass for the blue SDSS galaxies at 0.015$\leq~z~\leq$0.1. The separation between blue and red galaxies is defined in Fig.~\ref{FIG:SDSScolma}. The position of the Milky Way (MW) is plotted for comparison (M$_{\star}$(MW)=[6.1$\pm$0.5]$\times$10$^{11}$ M$_{\sun}$, Flynn et al. 2006; SFR(MW)=4 M$_{\sun}$ yr$^{-1}$, Diehl et al. 2006)).
              }
         \label{FIG:SFRMstar_SDSS}
   \end{figure}
%__________________________________________________________________
%__________________________________________________________________
\subsection{The SFR-stellar mass correlation}
\label{SEC:SFRMstar}
%__________________________________________________________________
%__________________________________________________________________
   \begin{figure}
   \centering
   \includegraphics[width=8cm]{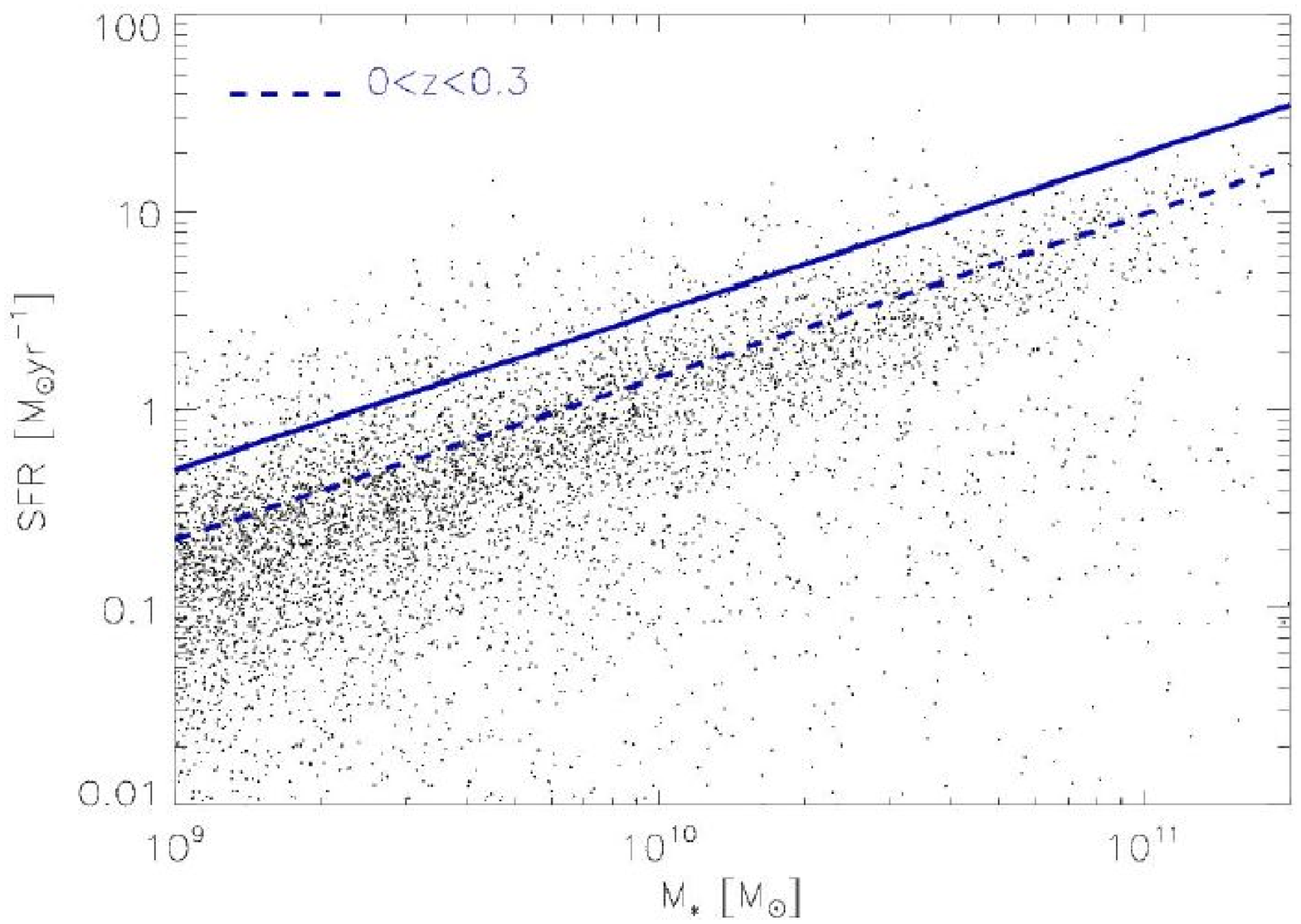}
    \includegraphics[width=8cm]{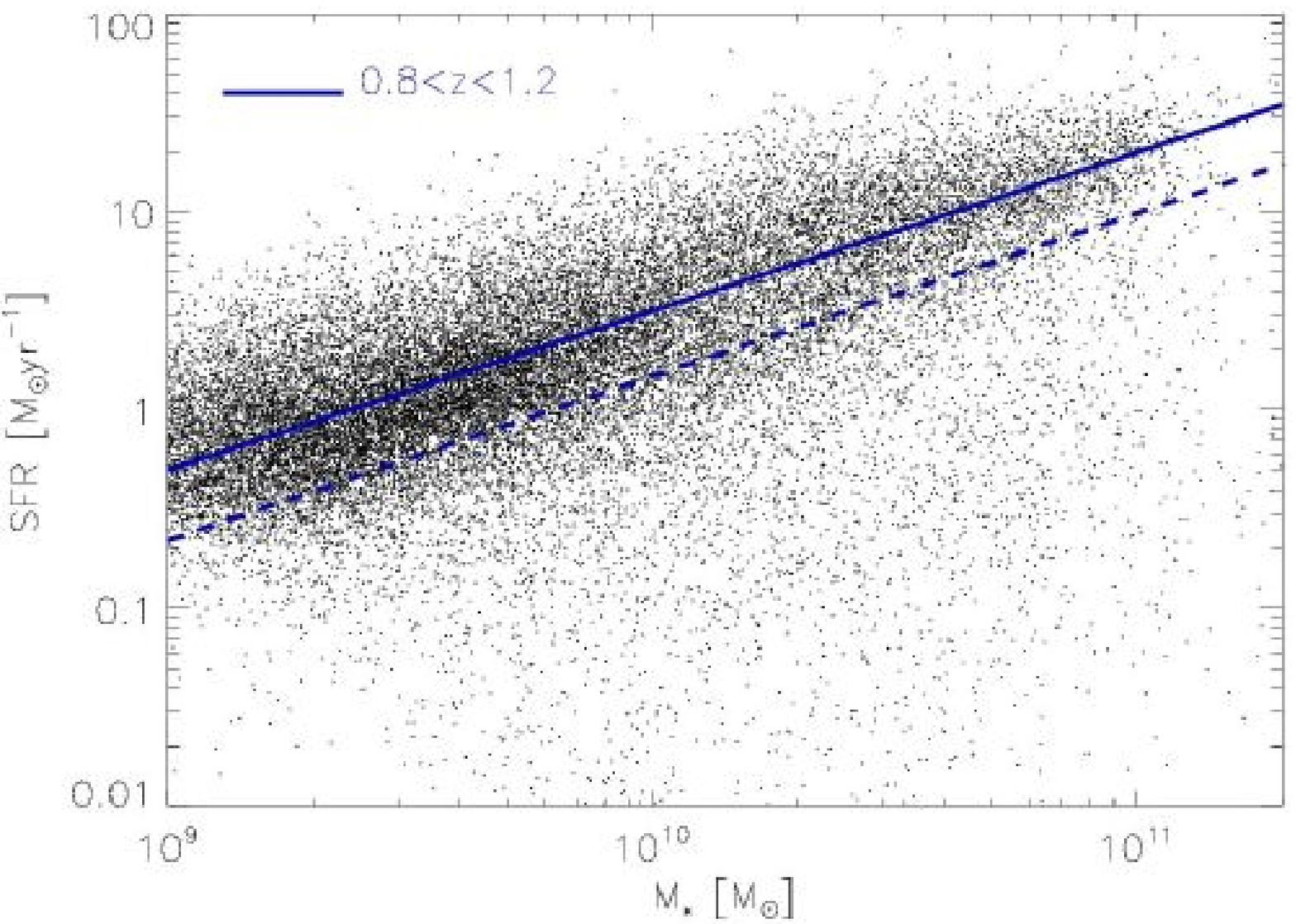}
     \caption{Correlation between the SFR and M$_{\star}$ predicted by the Millennium model for galaxies: (left) in the present-day universe ($z\leq$0.3, dashed line) ; (right) at $z$=0.8--1.2 (plain line). 
              }
         \label{FIG:SFRMstar_Mille}
   \end{figure}
%__________________________________________________________________
We find a tight correlation (0.3 dex rms) between the SFR and stellar mass of blue star-forming galaxies in GOODS at $z$=0.8-1.2 (Fig.~\ref{FIG:SFRMstar_GOODS}) fitted by Eq.~\ref{EQ:SFRM}. AGNs were excluded as in Fig.~\ref{FIG:mainfig}. This relationship is important to understand the origin of the reversal of the star formation-density relation. Indeed, if massive galaxies are preferentially located in dense regions and if the SFR increases with stellar mass, the physical reason why galaxies in dense environments are more rapidly forming stars at $z\sim$1 might not be directly related to the environment itself.

A similar correlation was found by Noeske et al. (2007a,b) in the Extended Groth Strip, although over a more limited range in SFR and masses. We see that galaxies are well separated in two clouds: one at high stellar masses and weak SFR and one exhibiting a continuous increase of SFR with M$_{\star}$. Note that red galaxies preferentially belong to the first cloud. However, note also that some blue galaxies fall in this cloud and that reversely some red galaxies fall in the correlation. Hence this correlation defines a more fundamental galaxy bimodality than that based on galaxy colors. 

The bottom part of Fig.~\ref{FIG:SFRMstar_GOODS} shows that the specific SFR (SSFR=SFR/M$_{\star}$) is slowly decreasing with stellar mass for $z\sim$1 galaxies, SSFR$\sim$M$_{\star}^{-0.1}$. The faster decrease seen in the first two bins is due to incompleteness below SFR= 3 M$_{\sun}$ yr$^{-1}$. Note that within the error bars, the SSFR is also consistent with a nearly constant trend between M$_{\star}$$\sim$10$^{10}$ and 10$^{11}$ M$_{\sun}$. The vertical error bars on SSFR are the bootstrapped 68\,\% confidence level on the measurement of the median SSFR value for each stellar mass bin. In the highest stellar mass bin, the number of galaxies is too small for bootstrapping, which underestimates the error bar, hence we used SSFR/sqrt(Number of galaxies) for this bin.

A similar correlation between the SFR and stellar mass has been shown for local galaxies with no AGN contribution in the SDSS (Fig.17 of Brinchmann et al. 2004). In this figure however, both red and blue galaxies are mixed together and the presence of a cloud of high-mass galaxies with low SFR can be mostly explained by the red galaxies (as defined in Fig.~\ref{FIG:GOODScolmag}). When taken separately, blue star forming galaxies exhibit a continuous increase of the SFR with stellar mass (see Fig.~\ref{FIG:SFRMstar_SDSS}) best-fitted by Eq.~\ref{EQ:sfrmstar_SDSS}. The slope of the local correlation ($\alpha$=0.77) is close but slightly smaller than at $z\sim$1 ($\alpha$=0.9). On average, the SFR of galaxies on the SFR-M$_{\star}$ correlation has increased by a factor six from $z\sim$0 to $z\sim$1 (see green line in Fig.~\ref{FIG:SFRMstar_GOODS}).

The Millennium model (see Fig.~\ref{FIG:SFRMstar_Mille}) overlaps very well with the SDSS at $z\sim$0 and predicts a slope very similar to the SDSS ($\alpha$=0.8) both at $z\sim$0 and 1 (see Eqs.~\ref{EQ:sfrmstarMillez0},\ref{EQ:sfrmstarMillez1}), but it also predicts an increase of the SFR by only a factor two, much below the observed trend (Fig.~\ref{FIG:SFRMstar_GOODS}).

 An important clue about the nature of distant LIRGs is given by the position of the galaxies M 82 and Arp 220 in the SFR--M$_{\star}$ diagram (Figs.~\ref{FIG:SFRMstar_GOODS},\ref{FIG:SFRMstar_SDSS}). A galaxy like the Milky Way exhibits a specific SFR slightly below the median trend for local galaxies but with the 68\,\% dispersion (Fig.~\ref{FIG:SFRMstar_SDSS}), but M82 is among the rare outlyiers with a SSFR about three times larger than the median, while the ULIRG Arp 220 exhibits a SSFR two orders of magnitude larger (see Fig.~\ref{FIG:SFRMstar_SDSS}). M82 is not a LIRG but its SED is very similar to the one of local LIRGs (Elbaz et al. 2002). If we now instead compare the position of those three galaxies to the $z\sim$1 trend, M82 becomes a typical galaxy for its mass, while Arp 220 still produces ten times more stars than an average galaxy of its mass and the Milky Way is ten times below the trend for its mass. It is of course unfair to compare local galaxies to $z\sim$1 ones, but it illustrates the fact that an actively star forming galaxy like M82 is a typical $z\sim$1 galaxy for its mass. In the following Section, we discuss the consequences of this correlation on the origin of the main result of this paper, the SFR-density relation.
%__________________________________________________________________
   \begin{figure}
   \centering
   \includegraphics[width=8cm]{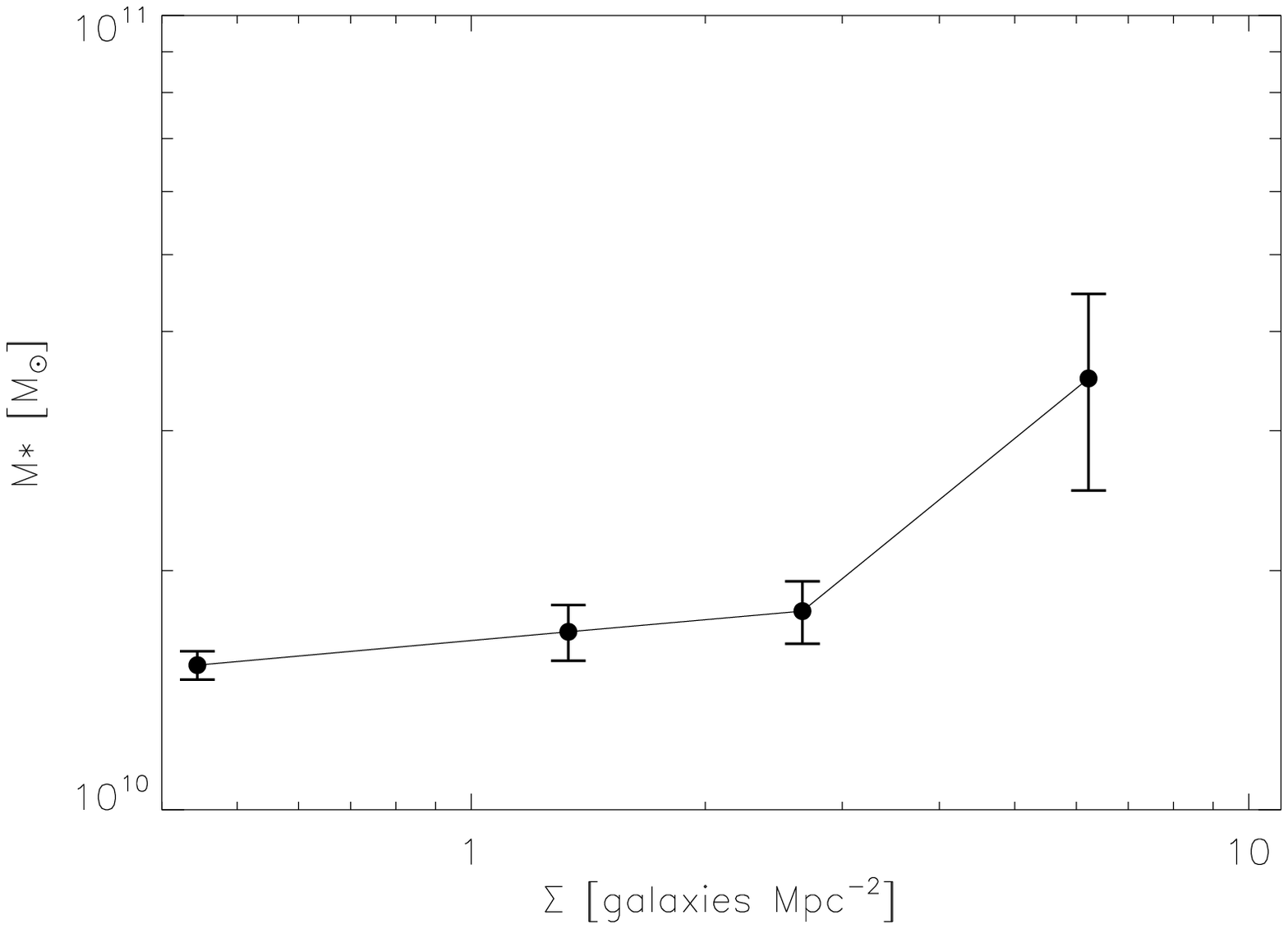}
   \includegraphics[width=8cm]{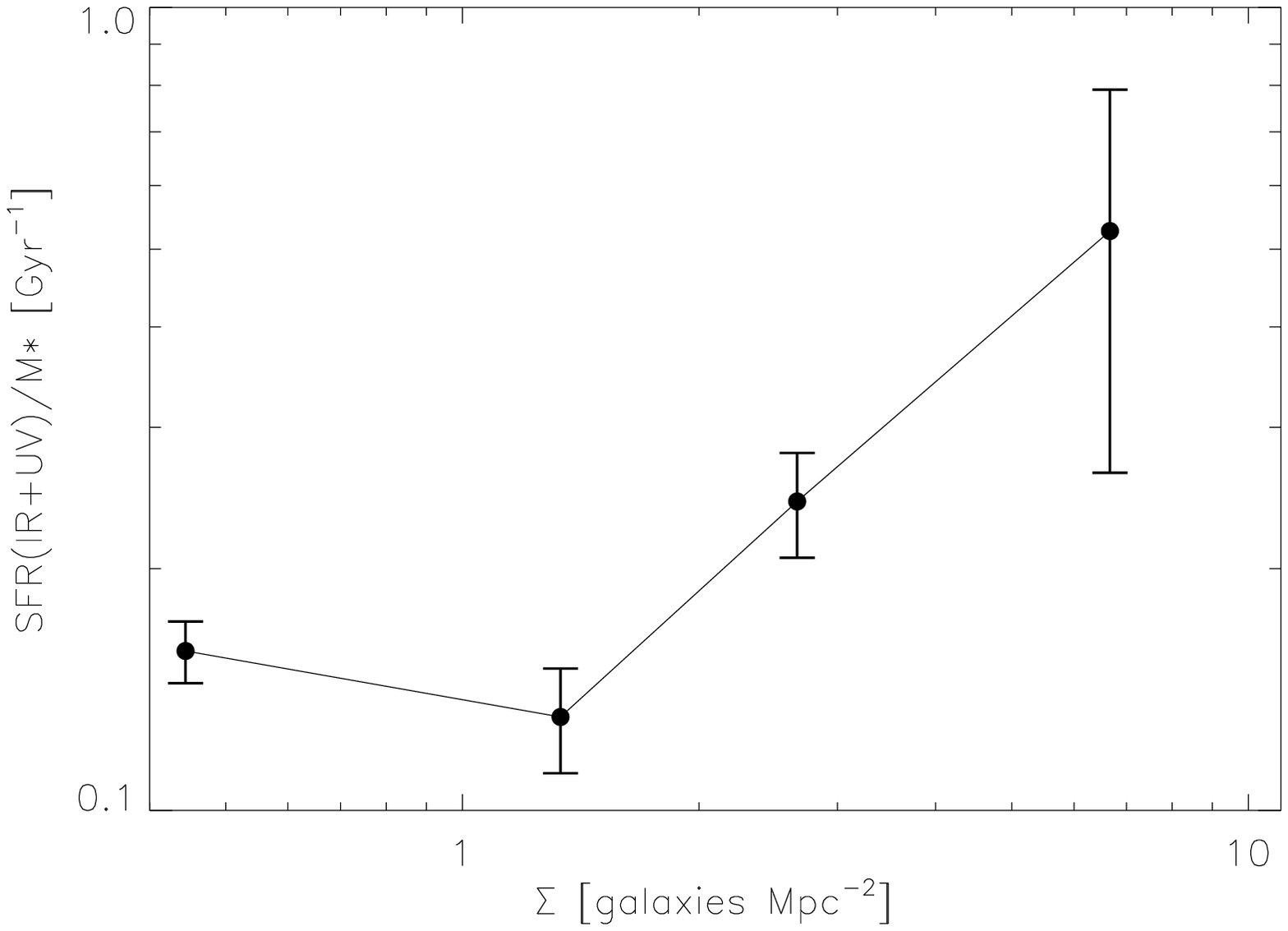}
      \caption{Evolution of the average stellar mass (M$_{\star}$; top) and specific star formation rate (SSFR=SFR/M$_{\star}$; bottom) of GOODS-N and GOODS-S galaxies as a function of galaxy density at z=0.8-1.2. SFR/M$_{\star}$ is only for galaxies with M$_{\star}$= 5$\times$10$^{10}$-5$\times$10$^{11}$ M$_{\sun}$, but $\Sigma$ is computed using all galaxies.
              }
         \label{FIG:Mstardens}
   \end{figure}
%__________________________________________________________________
%__________________________________________________________________
   \begin{figure*}
   \centering
   \includegraphics[width=13cm]{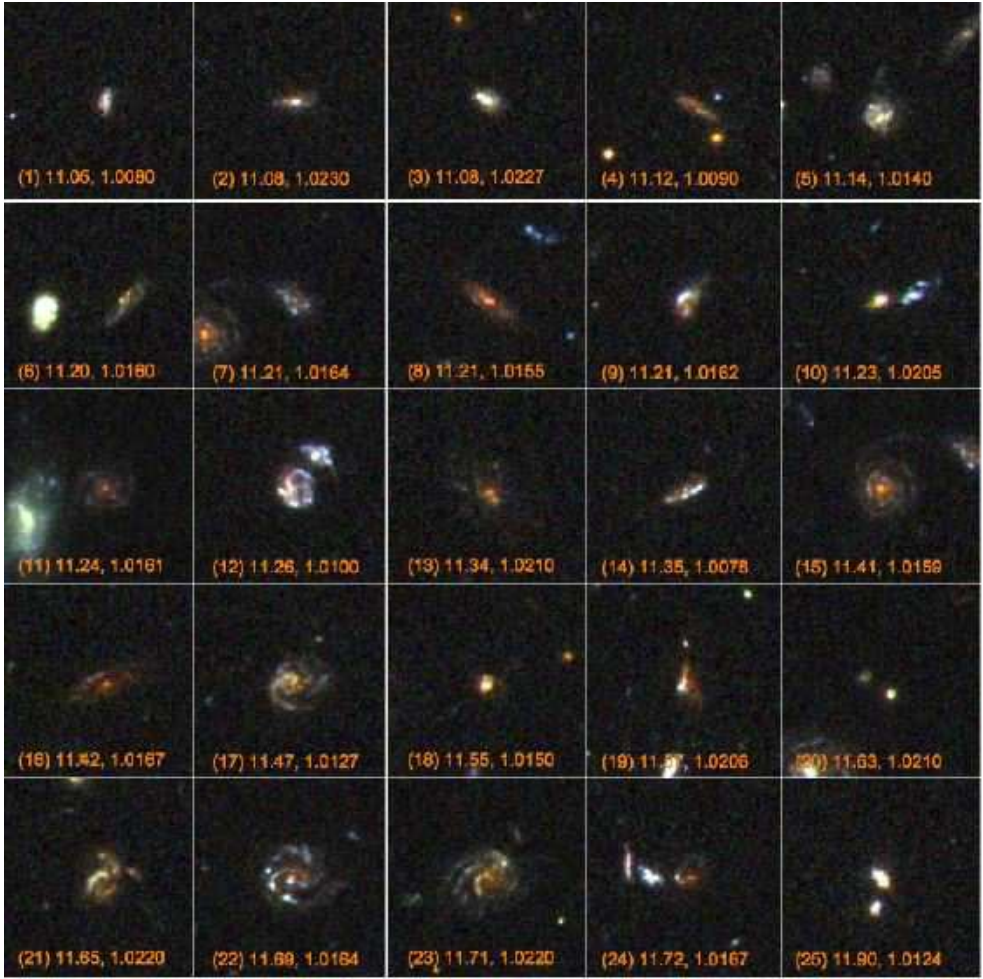}
      \caption{HST-ACS images of the 25 luminous infrared galaxies (LIRGS) in the GOODS-N structure at z=1.0163  (LIRGs: L$_{\rm IR}$=L(8-1000\,$\mu$m)$\geq$ 10$^{11}$ L$_{\sun}$, i.e. SFR $\geq$ 17 M$_{\sun}$yr$^{-1}$). The images result from the combination of the B, V and I passbands and their width is 25 kpc proper on a side (or 50 comoving-kpc, i.e. 3 arcsec at z=1.016). The labels show the ID between parenthesis, the logarithm of the total infrared luminosity in solar luminosities and the redshift. The galaxies were visually classified as major mergers (ID 6,7,9,10,12,13,15,19,24,25),  grand-design spirals (ID 5,8,11,16,17,21,22,23), compact galaxies (ID 1,2,3,18,20) and irregulars (ID 4,14).
              }
         \label{FIG:ACS}
   \end{figure*}
%__________________________________________________________________

\begin{equation}
\label{EQ:SFRM}
{\rm SFR}_{\rm GOODS}^{z\sim1} ~\left[ {\rm M}_{\sun}~ {\rm yr}^{-1}\right]=7.2~[-3.6,+7.2]\times \left[{\rm M}_{\star} / 10^{10} ~{\rm M}_{\sun}\right]^{0.9}
\end{equation}
\begin{equation}
\label{EQ:sfrmstar_SDSS}
{\rm SFR}_{\rm SDSS}^{z\sim0} ~\left[ {\rm M}_{\sun}~ {\rm yr}^{-1}\right]=8.7~[-3.7,+7.4]\times \left[{\rm M}_{\star} / 10^{11} ~{\rm M}_{\sun}\right]^{0.77}
\end{equation}
\begin{equation}
\label{EQ:sfrmstarMillez0}
{\rm SFR}_{\rm Millennium}^{z\sim0} ~\left[ {\rm M}_{\sun}~ {\rm yr}^{-1}\right]=1.5~[-0.9,+1.2]\times \left[{\rm M}_{\star} / 10^{10} ~{\rm M}_{\sun}\right]^{0.82}
\end{equation}
\begin{equation}
\label{EQ:sfrmstarMillez1}
{\rm SFR}_{\rm Millennium}^{z\sim1} ~\left[ {\rm M}_{\sun}~ {\rm yr}^{-1}\right]=3.2~[-1.8,+3.5]\times \left[{\rm M}_{\star} / 10^{10} ~{\rm M}_{\sun}\right]^{0.8}
\end{equation}

\subsection{Influence of the stellar mass and environment on the specific SFR at $z\sim$1}
\label{SEC:SSFR}
The increase of SFR with stellar mass at $z\sim$1 described in the previous section could provide a direct explanation for the SFR--$\Sigma$ relation if massive galaxies trace the regions of strong galaxy density. This is indeed the case although the effect is not very strong (Fig.~\ref{FIG:Mstardens}-left). In order to understand whether the SFR--$\Sigma$ relation is generated by the presence of massive galaxies in the densest regions, we have plotted the specific SFR (SSFR=SFR/M$_{\star}$) of massive galaxies, with M$_{\star}$= 5$\times$10$^{10}$-5$\times$10$^{11}$ M$_{\sun}$ (no bias in the mass selection as shown in Fig.~\ref{FIG:Mstarcompleteness}). By restricting ourselves to galaxies more massive than 5$\times$10$^{10}$ M$_{\sun}$, we remain in the domain where the sample is not affected by spectroscopic incompleteness. We find evidence that the specific SFR of massive galaxies does increase with galaxy density (Fig.~\ref{FIG:Mstardens}-right). The increase of both M$_{\star}$ and SFR/M$_{\star}$ with $\Sigma$ are therefore together responsible for the trend of SFR with $\Sigma$ seen in Fig.~\ref{FIG:mainfig}. But the fact that SFR/M$_{\star}$ decreases with M$_{\star}$ (see Fig.~\ref{FIG:SFRMstar_GOODS}) while it increases with $\Sigma$ favors the interpretation that the environment does enhance the SFR activity of galaxies. 
\subsection{Influence of the morphology on the specific SFR of LIRGs at $z\sim$1}
\label{SEC:mergers}
While LIRGs make a negligible fraction of local galaxies in number and produce less than 2\,\% of the bolometric luminosity per unit volume in the present-day Universe, they are the galaxies which exhibit the fastest evolution between $z$=0 and 1, with an increase by a factor 70 of their luminosity per unit comoving volume (Chary \& Elbaz 2001, Le Floc'h et al. 2005). Consequently, the average SFR of a galaxy at $z\sim$1 is that of a LIRG (Elbaz et al. 2002). They play a major role in the reversal of the star formation--density relation at $z\sim$1 presented here and are responsible for the peak of $<$SFR$>$ in dense regions in Fig.~\ref{FIG:mainfig}. Hence in order to understand the origin of the reversal, it is also necessary to understand the origin of the evolution of LIRGs with redshifts.

In the present-day Universe, the vast majority of LIRGs are triggered by the major merger of two massive spiral galaxies (mass ratio larger than 1/4), which can be identified either by a disturbed morphology with tidal tails or by the presence of a companion massive spiral within 30 kpc (Ishida 2004). There are 221 LIRGs (140 in GOODS-N and 81 in GOODS-S) with a spectroscopic redshift within 0.8$\leq z \leq$1.2 in the GOODS fields after excluding AGNs. The fraction of galaxies which harbor an AGN (identified as in Bauer et al. 2004) that we would have classified as LIRGs from their observed 24\,$\mu$m flux density alone, hence for which the true IR luminosity and SFR are uncertain, is 22\,\% in GOODS-N and 16\,\% in GOODS-S.

We visually divided LIRGs between four morphological types from their composite B,V,I HST-ACS images, as illustrated by the HST-ACS images of the 25 LIRGs located in the structure at $z\sim$1.016 in Fig.~\ref{FIG:ACS}. We find that 46\,\% of the LIRGs at $z\sim$1 exhibit the morphology of grand design spirals, while only 31\,\% are showing evidence for a major merger. The remaining galaxies are divided into compact galaxies (9\,\%) and irregulars (14\,\%). Hence only a third of the LIRGs at $z\sim$1 present direct evidence suggesting that they are experiencing a major merger (see also Bell et al. 2005, Zheng et al. 2004, Melbourne et al. 2006). The rapid decline of the cosmic SFR density since $z\sim$1, which is for a large part due to LIRGs (see Le Floc'h et al. 2005), is therefore probably not caused by the progressive disappearance of major mergers (see also Lotz et al. 2007). It is more probably caused by the exhaustion of the gas available for star formation by the galaxies. We note however, that some galaxies that we classified as major mergers might not be such, since even in the deep ACS images in GOODS, tidal tails can hardly be observed. Reversely, some fraction of the irregulars, which are the most difficult to classify could be associated with a major merger as well as some compact galaxies, which could well correspond to more advanced stages in the merging process (e.g. ID 1 in Fig.~\ref{FIG:ACS}). Hence the true fraction of major mergers can vary between $\sim$30 and 50\,\% of the galaxies. Here we are only discussing the role of major mergers, because this is the only mechanism that we can hope to visually identify in the $z\sim$1 LIRG population, but there are many other ways through which the environment can act on galaxies such as minor mergers (mass ratio $<$1/4), galaxy harassment during group formation, dynamical instabilities during tidal encounters (Combes 2005) or even the acceleration of intergalactic gas infall during group formation. 

\begin{figure}
   \centering
   \includegraphics[width=8cm]{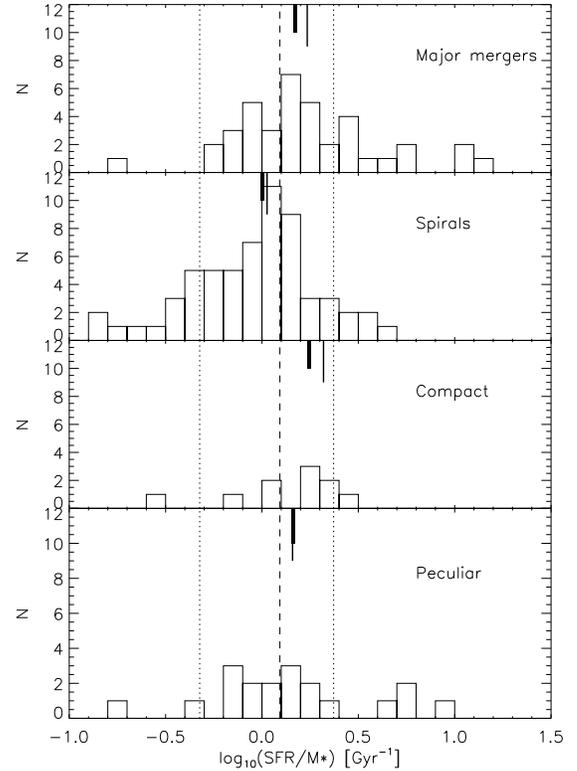}
      \caption{Distribution of the specific SFR (SFR/M$_{\star}$) of LIRGs in GOODS-N within 0.8$\leq z \leq$1.2 for the four morphology classes: major mergers, spirals, compact and irregulars. The thick and thin lines on the top of each figure indicate the median and average specific SFR respectively, for each sample. The dashed line shows the median specific SFR for the whole sample of $z\sim$1 LIRGs and the surrounding dotted lines indicate the region including 68\,\% of the galaxies.             }
         \label{FIG:his_morpho}
   \end{figure}

In Fig.~\ref{FIG:his_morpho}, we compare the distribution of the specific SFR for the four morphology classes among $z\sim$1 LIRGs. Unexpectedly, the majority of the galaxies of the four classes are distributed inside the interval centered on the median specific SFR and including 68\,\% of the $z\sim$1 LIRGs, including even the galaxies presenting the morphological signature of an on-going major merger. Hence the event of a major merger does not seem to produce a major impact on the specific SFR of a galaxy. This may explain why the relation between the SFR and stellar mass (Sect.~\ref{SEC:SFRMstar} and Fig.~\ref{FIG:SFRMstar_GOODS}) presents only a 0.3 dex dispersion. Indeed if mergers were inducing a major change in the star formation activity of a LIRG at $z\sim$1, the dispersion would be larger as discussed in Noeske et al. (2007b).
We note however the presence of a tail on the high specific SFR side for major mergers and on the opposite side for Spirals. The median and mean specific SFR of major mergers are 50 and 70\,\% larger than those of Spirals. The time it would take for a $z\sim$1 LIRG with a spiral morphology to double its stellar mass at continuous SFR is $\sim$ 1 Gyr, while it is $\sim$ 0.6 Gyr for a LIRG experiencing a major merger. Finally, we note that the LIRGs with a compact morphology exhibit the largest average specific SFR, equivalent to a doubling timescale of $\sim$ 0.5 Gyr. This favors the hypothesis that they advanced stages of major mergers.
%__________________________________________________________________
%__________________________________________________________________

\section{Conclusions}
We have presented evidence that the SFR of individual galaxies was increasing with the environment up to a critical galaxy density above which it decreases again. This reversed trend, with respect to the one observed locally, suggests that galaxy evolution is not independent from structure formation at larger scales in the universe. The trend is present in both GOODS fields, hence it is robust against cosmic variance. This result is reinforced by the finding that $z\sim$1 LIRGs present a large correlation length typical of local L$_{\star}$ ellipticals (Gilli et al. 2007).

A large part of this paper was dedicated to the validation of this result against selection effects both in spectroscopic redshifts, stellar mass and SFR estimates, as well as against the presence of active galactic nuclei. This result was discussed in the framework of the bimodality of galaxies, separated between red-dead and blue-star forming galaxies. We demonstrated that the combination of IR and UV measurements are crucial to detect such trend and that even the UV corrected for extinction alone was insufficient due to a saturation around 10 M$_{\sun}$ yr$^{-1}$.

The role of structure formation is exemplified by a zoom on a structure at $z$=1.016, where galaxies are clearly distributed in groups, with the most strongly star forming ones at the center of these groups. This both illustrates that it is the individual SFR of galaxies that increases with density (not the sum of all SFR per unit volume or area) and that the typical scale that affects star formation in galaxies at $z\sim$1 is the $\sim$1--2 Mpc one, interestingly consistent with local observations (Blanton et al. 2006). 

We have discussed the possibility that the SFR--density relation at $z\sim$1 could potentially be due to the presence of larger stellar masses in denser environments and not necessarily to direct environmental effects. We showed that there is indeed a correlation between the SFR and stellar mass of galaxies both at $z\sim$0 (SDSS) and $z\sim$1 (GOODS), as also found over a smaller range of SFR and stellar masses by Noeske et al. (2007a,b). This suggests that the increase of the stellar mass with increasing galaxy density is responsible for part of the SFR--density trend at $z\sim$1. 

However, the specific SFR (=SFR/M$_{\star}$) was found to decrease with stellar mass while it increases with galaxy density within a limited stellar mass range (M$_{\star}\sim$5$\times$10$^{10}$ -- 5$\times$10$^{11}$ M$_{\sun}$). The error bars on these trends are important because of the small size of GOODS, but they favor the interpretation that the environment does affect the star formation activity of galaxies.

In order to understand how the environment affects the activity of galaxies, we visually classified the most actively star forming ones (LIRGs at 0.8$\leq z \leq$1.2) in four morphology classes (major mergers, grand design spirals, compact and irregular galaxies). We found that about $\sim$46\,\% of them exhibit the morphology of grand design spirals, while the fraction of galaxies showing evidence for a major merger is $\sim$31\,\%. The remaining galaxies are divided into compact galaxies (9\,\%) and irregulars (14\,\%). Hence major mergers alone cannot explain the rapid decline of the cosmic SFR density since $z\sim$1, and in particular that of the LIRG contribution, dominant at $z\sim$1 and negligible at $z\sim$0. 

The average specific SFR of major mergers is 70\,\% larger than that of spirals. Nearly all galaxies in the low specific SFR tail of the distribution are spirals, while very few spirals belong to the high specific SFR tail. However most galaxies of both morphology classes lie around the same value at the 68\,\% confidence level. Hence major mergers do not appear to strongly affect the efficiency with which galaxies are forming stars and therefore appear not to be the major cause for the reversal of the star formation--density relation at $z\sim$1. Several other mechanisms of galaxy interactions could play a role that remain to be identified, such as minor mergers (mass ratio $<$1/4), galaxy harassment during group formation, dynamical instabilities during tidal encounters (Combes 2005) or even the acceleration of intergalactic gas infall during group formation. 

To conclude, we are witnessing at $z\sim$1 the continuing, vigorous formation of massive galaxies in connection with structure formation, a process that will be definitively completed only at later epochs.  While we do see that merging activity is taking place, it appears that in situ formation of stars at large rates is also an important channel for mass growth in galaxies. Reproducing the SFR--density relation at $z\sim$1 is a new challenge for models, and provides a crucial test of the correct balance between mass assembly through these two channels at early epochs.

\begin{acknowledgements}
We wish to thank Jarle Brinchmann and Alexis Finoguenov for sharing their work with us. D.Elbaz and D.Le Borgne wish to thank the Centre National dÕEtudes Spatiales for his support. D.Elbaz thanks the Spitzer Science Center at the Caltech university for support. D. Alexander thanks the Royal Society for support.
\end{acknowledgements}

\end{document}